\documentclass[%
 12pt,
 tightenlines,
 superscriptaddress,
 nofootinbib,
 notitlepage,
 bibnotes,
 amsmath,amssymb,
 aps,
 prd,
]{revtex4-1}
\usepackage{units}
\usepackage{graphicx}
\usepackage{dcolumn}
\usepackage{bm}
\usepackage{leftidx}
\usepackage{fouriernc}
\usepackage{amsfonts,amsmath,epsfig,bm,physics}
\usepackage[dvipsnames]{xcolor}
\usepackage{graphicx,hyperref}
\usepackage{booktabs}
\usepackage{esvect}
\usepackage[caption=false]{subfig}
\hypersetup{
    colorlinks,
    citecolor=red,
    filecolor=red,
    linkcolor=Maroon,
    urlcolor=Maroon
}

\begin{document}

\preprint{APS/123-QED}

\title{Primordial black hole formation in inflationary $\alpha$-attractor models}

\author{Rafid Mahbub}
\email{mahbu004@umn.edu}
\affiliation{%
 School of Physics \& Astronomy, University of Minnesota,\\
 Minneapolis, MN 55455, USA 
}%

\date{\today}

\begin{abstract}
The formation of primordial black holes is studied using superconformal inflationary $\alpha$-attractors. An inflaton potential is constructed with a plateau-like region which brings about the onset of an ultra slow-roll region where the required enhancement in the curvature power spectrum takes place. This is accomplished by carefully fine-tuning the parameters in the potential. For the parameter sets, the amount of observable inflation is such that the curvature perturbation $\mathcal{P}_{\mathcal{R}}$ peaks at $\sim\unit[10^{14}]{Mpc^{-1}}$, producing PBHs in the mass window $\unit[10^{16}]{g}\leq M \leq \unit[10^{18}]{g}$. The reheating period after inflation is taken into account to determine whether or not PBH formation takes place in such a phase. Finally, the spectrum of second order gravitational waves, that can arise due to the curvature perturbation enhancement, is calculated.
\end{abstract}

\maketitle


\section{Introduction}
\begin{figure}[t]
\centering
\includegraphics[scale=0.6]{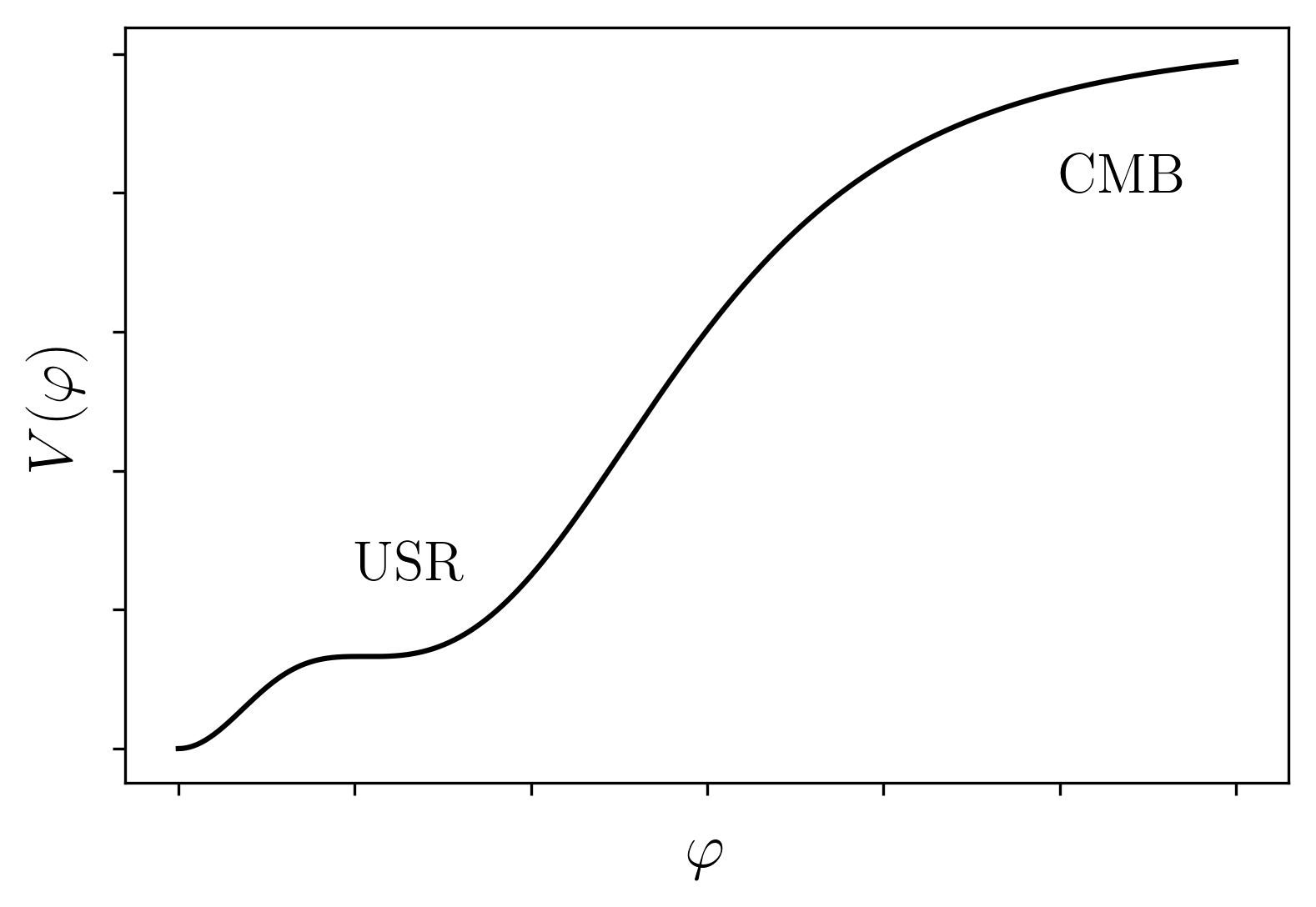}
\includegraphics[scale=0.6]{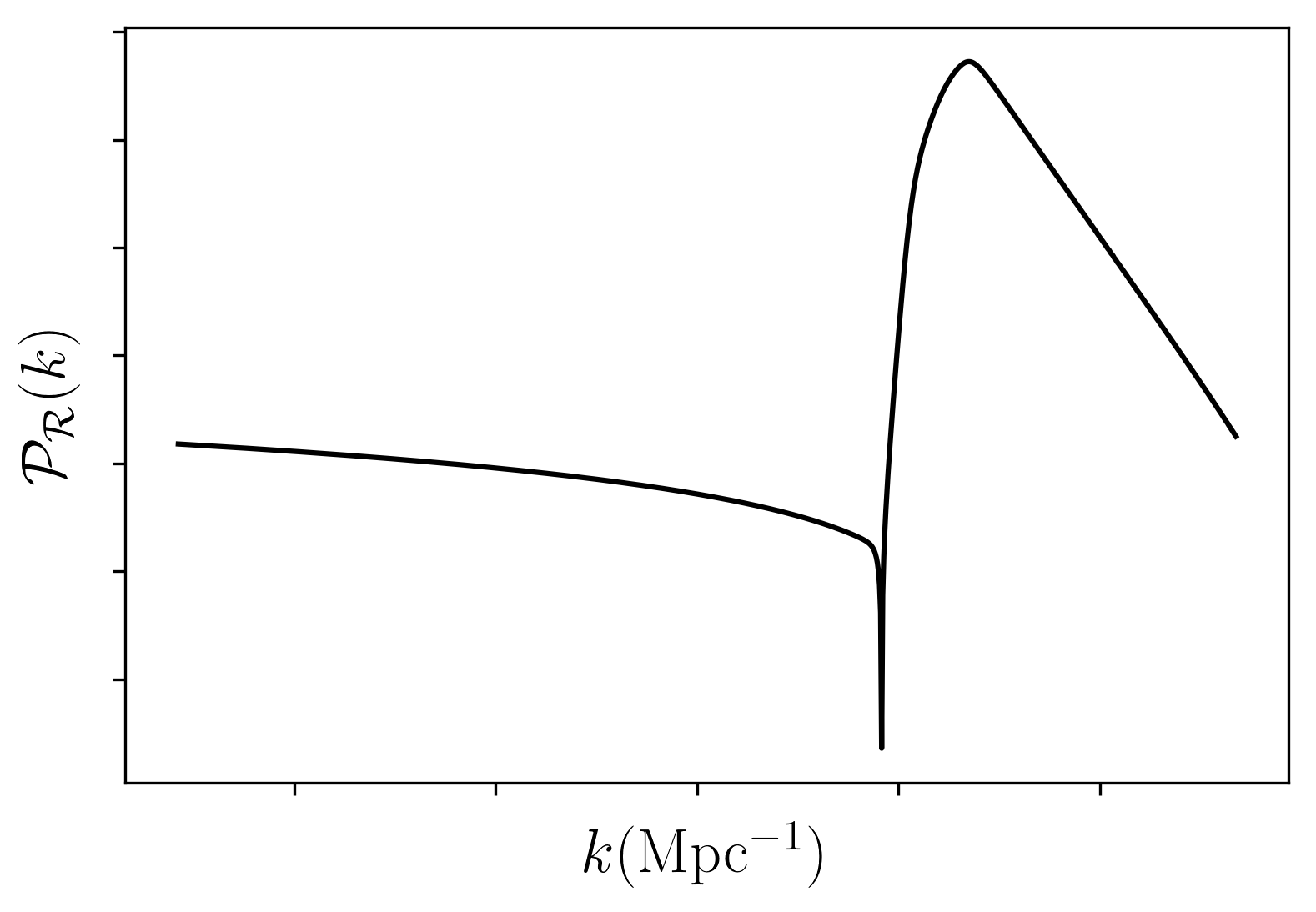}
\caption{Sketch of the kind of inflaton potential (left) that gives rise to small scale enhancement of the curvature power spectrum (right). The curvature perturbation peaks at $10^{-2}$ for modes $k_{\text{peak}}\sim \unit[10^{14}]{Mpc^{-1}}$}.
\label{fig:sketch}
\end{figure}

The collapse of overdense regions in the radiation plasma of the very early universe, forming black holes, was an idea that was first proposed over forty years ago by Zeldovic and Novikov \cite{zeldovich}: later being explored by Hawking and Carr in the 70's, producing a series of seminal papers on the subject \cite{carr1,hawking2,hawking}. Although there has been continuous work on the subject over the decades, interest in these primordial black holes (PBHs) was renewed after the first gravitational wave (GW) discovery, GW150914, from the coalescence of two $\sim30M_{\odot}$ black holes \cite{abbott,sasaki2}.\\
\indent PBHs are different from astrophysical black holes in that they do not form from the gravitational collapse of stars. Hence they are termed as primordial: being formed during the radiation-dominated phase of the early universe. Due to their non-stellar origins, their masses are not bound by the Chandrasekhar limit and a detection of black holes with mass $M< M_{\odot}$ would be an ideal signature of primordial origins. These black holes are supposed to have formed from the gravitational collapse of rare primordial density perturbations that are of the order unity upon horizon entry. During the 90's and early 2000's, research on PBH formation was done mainly using QCD phase transitions since it was believed, at the time, that the QCD phase transition in the early universe was first order \cite{jedamzik1997}. Such a phase transition would have produced a fluid with zero pressure gradient, conducive to gravitational collapse. However, it is now known that the QCD phase transition is actually a crossover for low baryon chemical potential $\mu_{\text{B}}$ which was the case during the early universe \cite{bellido}. As a result, inflation has become the prime candidate that is used to study PBH formation. Since inflation already explains how density perturbations are produced, which later lead to structure formation, it is naturally suited for such explorations.\\
\indent However, PBH production during the radiation epoch implies the collapse of rare overdense regions of the order unity and such rare peaks in the density field cannot be produced using conventional slow-roll models of inflation which predict a nearly scale invariant curvature power spectrum of $\mathcal{P}_{\mathcal{R}}\sim 10^{-9}$ at CMB scales, decreasing only slowly with scale. The density perturbations in the post-inflationary stages are proportional to the comoving curvature perturbations generated during inflation and $\mathcal{O}(1)$ density perturbations require an enhancement in the curvature perturbation power spectrum, $\mathcal{P}_{\mathcal{R}}$, of around $10^{-2}$ at small scales.\footnote{These are the scales that re-enter the horizon very soon after the end of inflation, giving rise to PBHs of lower mass.} Such an amplification is possible in models which exhibit certain peculiarities in the inflaton potential. Amplification of the curvature power spectrum has been shown to be achievable in various models such as multi-field inflation models like hybrid inflation \cite{garcia1996density,bellido1}, double inflation \cite{kannike2017single} and a collection of models where the inflaton potential contains a plateau-like region exhibiting ultra slow-roll (USR) dynamics such as (i) string theory inspired models \cite{ozsoy,Cicoli:2018asa} (ii) critical Higgs inflation \cite{bellido3} (iii) radiative plateau inflation \cite{ballesteros} (iv) MSSM inflation \cite{Choudhury:2011jt} and (v) inflationary $\alpha$-attractors \cite{dalianis} among others.\\
\indent There has been much work done recently on inflation models with a \textit{quasi}-inflection point in the potential near the end of inflation. Such a feature introduces a USR period where the inflaton slows down and the necessary amplification of the curvature power spectrum takes place \cite{dimo,pattison}. The scales which exit the horizon during this time later re-enter shortly after the end of inflation, whereby the radiation plasma inherits the high curvature perturbation as an increase in the overdensity. To understand how such an increase takes place, the analytic form of the curvature perturbation can be considered (under the slow-roll approximation), which is $\mathcal{P}_{\mathcal{R}}\sim H^{2}/\epsilon$, where $\epsilon$ is the first Hubble flow parameter. It so happens that, during USR, $\epsilon$ falls very rapidly and one can be convinced that $\mathcal{P}_{\mathcal{R}}$ will experience an increase. One way to understand the rapid decrease in $\epsilon$ is through its equivalence with the potential slow-roll parameters $\epsilon\simeq\epsilon_{\text{V}}$, where $\epsilon_{\text{V}}$ is proportional to $\partial_{\phi}V$. This implies that around a plateau where $\partial_{\phi}V\sim 0$, $\epsilon_{\text{V}}$ will become greatly reduced. Nevertheless, without careful fine-tuning, in the vicinity of a region where $\partial_{\phi\phi}V=0$ and $\partial_{\phi}V\approx0$, the inflaton may simply slow down without the required increase in the power spectrum. The slow-roll formula is ill-suited to correctly enumerate the curvature power spectrum since it is also derived under the assumption that the Hubble parameter does not change appreciably during inflation. In fact, along the plateau, inflation breaks away from the quasi-de Sitter expansion and there is a significant change in the Hubble parameter. Hence, an accurate computation of the curvature power spectrum requires solving the Mukhanov-Sasaki (MS) equation which describes how the Fourier modes of the curvature perturbations evolve from an initially subhorizon state to a superhorizon one. Numerical solutions of the MS equation reveal that the analytic expression underestimates the curvature power spectrum by at least one order of magnitude. A sketch of the form of the potential and curvature power spectrum is shown in Fig. \hyperref[fig:sketch]{1}.\\
\indent The nature of dark matter has been one of the most important outstanding questions in physics. There is ample evidence of its existence. Yet we do not know what it is composed of. There are more popular particle dark matter candidates: axions, weakly interacting massive particles (WIMPs) and supersymmetric candidates (eg. neutralinos) which have not been observed yet. It is because of this that PBHs have become an attractive alternative for a dark matter candidate \cite{clesse2018,carr2}. If they exist, they are likely to have been formed before the Big Bang Nucleosynthesis (BBN) and escape the restrictions imposed on baryonic dark matter \cite{carr2016constraints}. They are also collisionless and non-relativistic - properties which are characteristic of cold dark matter (CDM). Nevertheless, there are certain restrictions on the types of inflation models that can account for dark matter this way. Single field inflation models with quasi-inflection points require a careful balance between the parameters and, quite often, a considerable amount of fine-tuning to ensure that the inflaton slows down enough so that the curvature power spectrum gets amplified by the correct proportion. Also, the parameters need to be adjusted in such a way so as to avoid a scenario where the inflaton gets stuck in the plateau, giving rise to an unnecessarily large number of e-folds. Observational evidence shows us that PBHs cannot comprise the totality of DM since the fraction of PBH energy density over CDM is constrained over a wide mass range. Nevertheless, there are a couple of mass windows that are rather weakly constrained and well tailored models can produce PBH that constitute around 10\% of the dark matter.\footnote{The constraints are more severe for some mass ranges than others, eg. $M\sim10^{15}-\unit[10^{16}]{g}$ is severely constrained by a lack of extragalactic $\gamma$-ray background resulting from evaporating black holes. On the other end of the spectrum, massive PBHs are also largely constrained by the CMB.}\\
\indent In this paper, the formation of PBHs is studied by constructing a potential using superconformal inflationary $\alpha$-attractors along the lines of \cite{dalianis}. The potential considered here has been influenced from a type of deformed Starobinsky potential that can be fine-tuned to produced a plateau where a USR period enhances the curvature perturbations, peaking at around $k\sim \unit[10^{14}]{Mpc^{-1}}$. The Mukhanov-Sasaki equation is numerically solved to evaluate the curvature power spectrum for all the Fourier modes of interest. Then, applying the Press-Schechter formalism, the fraction of energy density of PBH in DM is calculated (ignoring evaporation and accretion effects). Although conventional $\alpha$-attractor models predict a scalar spectra index of $n_{s}=1-2/N$, where $N$ is the amount of observable inflation, it is found that the value of $n_{s}$ predicted by these models is somewhat lower, where a better estimate of the spectral index is given by $n_{s}\approx 1-2/N_{\text{peak}}$. Here, $N_{\text{peak}}$ is the amount of e-folds measured between the points when the observable scale $k_{\text{CMB}}^{-1}$ and $k_{\text{peak}}^{-1}$ exit the horizon. In this work, two parameter sets have been considered. Each parameter set is further divided into two sets of observable e-folds in order to compute the curvature power spectra. PBHs produced by this model have masses in the range $\unit[10^{16}]{g}\leq M \leq \unit[10^{18}]{g}$ since the modes which collapse are ones which re-enter the horizon shortly after inflation. This also prompts one to consider whether or not these parameter sets predict a prolonged duration of reheating after inflation. In such a case, it could be possible that the relevant, PBH forming scales re-enter the horizon during reheating. The work is concluded by evaluating the energy spectrum of second order gravitational waves that can arise from such a model.\\
\indent In this work the natural units are used, where $c=\hbar=1$ and also setting $M_{p}=(8\pi G)^{-1}$ to unity unless otherwise stated. For the metric tensor, the mostly positive convention $(-+++)$ on an Friedmann-Lemaitr\'{e}-Robertson-Walker (FLRW) background is used, where $ds^{2}=-dt^{2}+a^{2}(t)d\bm{x}^{2}=a^{2}(\tau)(-d\tau^{2}+d\bm{x}^2)$ in cosmic and conformal times. Derivatives with respect to conformal time are represented with primes $(...)'$.
\section{Setup of the inflaton potential}
To achieve the kind of inflaton dynamics that is sought after, a potential needs to be constructed which allows for a large period of slow-roll, around the time observable scales exit the horizon, followed by a plateau where USR can take place. For such a potential, the usual slow-roll approximations should be valid up to the point where the inflaton starts to enter the USR phase. It has been suggested in \cite{ketov2018} that a modification of the Starobinsky potential, by adding an extra exponential term quadratically dependent on $\phi$, can help achieve this. The potential has the following form
\begin{equation}
V(\phi)=V_{0}\left( 1 + a_{1} - e^{-a_{2}\phi}-a_{1}e^{-a_{3}\phi^2} \right)^2
\end{equation}
where $a_{2}=\sqrt{2/3}$. The potential assumes the form of Starobinsky in the limit $a_{1}\longrightarrow 0$. This potential has the feature that it becomes flat for asymptotically large values of $\phi$ and the USR plateau can be placed in a desired location by fine-tuning the parameters. By demanding the existence of an inflection point, $\tilde{\phi}$, constraints on the parameters $a_{1},a_{3}$ can be imposed-
\begin{equation}\label{eq:constraints}
a_{1}=-\frac{a_{2}}{2a_{3}\tilde{\phi}}e^{-a_{2}\tilde{\phi}+a_{3}\tilde{\phi}^2}\;\;\;\;\;\;\;a_{3}=\frac{a_{2}\tilde{\phi}+1}{2\tilde{\phi}^2}
\end{equation}
The other parameter $V_{0}$ is set by imposing CMB constraints on the primordial power spectrum at the observable scale. At this point, $V(\phi)$ looks like a toy model and perhaps it could be incorporated into a more fundamental framework like supergravity. In the following section, this potential shall be adapted into the superconformal $\alpha$-attractor model of inflation. One of the benefits of using the $\alpha$-attractors is that, after stabilizing the inflaton trajectory, the potential takes the form of an arbitrary holomorphic function in terms of $\phi$.
\subsection{Superconformal $\alpha$-attractor inflation}
Let us consider an $\mathcal{N}=1$ superconformal theory coupled to chiral multiplets $X^{I}$, with $I=(0,1,...,n)$. These superconformal theories have an additional chiral superfield $X^0$, called the conformal compensator (or simply 'conformon'). The superconformal theory is specified by the functions $N(X,\bar{X})$ and $\mathcal{W}(X)$ with which the Lagrangian is constructed \cite{kallosh,kallosh2}
\begin{equation}
\frac{1}{\sqrt{-g}}\mathcal{L}=-\frac{1}{6}N(X,\bar{X})R-G_{I\bar{J}}\mathcal{D}^{\mu}X^{I}\mathcal{D}_{\mu}\bar{X}^{\bar{J}}-G^{I\bar{J}}\mathcal{W}_{I}\bar{\mathcal{W}}_{\bar{J}}
\end{equation}
where $N(X,\bar{X})$ is the K\"{a}hler manifold of the embedding space, $\mathcal{W}$ is the superpotential,$G_{I\bar{J}}=\partial_{I\bar{J}}N$ and $\mathcal{D}_{\mu}X^{I}=\partial_{\mu}X^{I}-iA_{\mu}X^{I}$.\footnote{These theories require a non-minimal coupling of the complex scalars to the curvature.} The superconformal theory has extra symmetries, apart from local supersymmetry, such as Weyl symmetry, special conformal symmetry, $U(1)_{\text{R}}$ symmetry and special supersymmetry. The existence of these additional symmetries requires the presence of the conformon \cite{kallosh3}. The goal is then to derive Poincar\'{e} supergravity from the underlying theory. Poincar\'{e} supergravity has the following Lagrangian
\begin{equation}
\frac{1}{\sqrt{-g}}\mathcal{L}=\frac{1}{2}R-K^{\Phi\bar{\Phi}}\partial_{\mu}\Phi\partial^{\mu}\bar{\Phi}-V(\Phi,\bar{\Phi})
\end{equation}
where the F-term scalar potential is
\begin{equation}\label{eq:F_pot}
V=e^{K}\left[ D_{\Phi_{i}}WK^{\Phi_{i}\bar{\Phi}_{j}}D_{\bar{\Phi}_{j}}\bar{W}-3|W|^2 \right]
\end{equation}
and where 
\begin{equation}
D_{\Phi_{i}}W=\frac{\partial W}{\partial\Phi_{i}}+\frac{\partial K}{\partial\Phi_{i}}W
\end{equation}
is the K\"{a}hler covariant derivative. In order to derive Poincar\'{e} supergravity, we consider the minimal chiral multiplet useful for cosmology which, other than the conformon $X^{0}$. contains two chiral superfields. They are $X^{1}=\Phi$ (which will later be identified with the inflaton) and $X^{2}=S$ called the stabilizer (or the sGoldstino). There is a large universality class of models associated with the superconformal theory.  We consider one with the following K\"{a}hler and superpotential for the embedding manifold: 
\begin{align}
N(X,\bar{X})&=-|X^0|^{2}\left( 1-\frac{|X^1|^{2}+|S|^2}{|X^0|^2} \right)^{\alpha}\\
\mathcal{W}(X)&=Sf\left( \frac{X^1}{X^0} \right)(X^0)^2\left( 1-\frac{(X^1)^2}{(X^0)^2} \right)^{(3\alpha-1)/2}
\end{align}
The $N(X,\bar{X})$ is only manifestly symmetric under $SU(1,1)$ when $\alpha=1$ and, for constant $f$ and $\alpha=1$, the superpotential preserves the subgroup $SO(1,1)$. However, to ensure that none of the fields become tachyonic, an extra term is added to the K\"{a}hler such that 
\begin{equation}
N(X,\bar{X})=-|X^0|^{2}\left( 1-\frac{|X^1|^{2}+|S|^2}{|X^0|^2} +3g\frac{|S|^4}{|X^0|^2(|X^0|^2 -|X^1|^2)}\right)^{\alpha}
\end{equation}
With these choices, gauge fixing is performed on the conformon - in particular using the D-gauge where $X^{0}=\bar{X}^{0}=\sqrt{3}$. Gauge-fixing of the conformon ensures that superconformal symmetry is spontaneously broken, producing a supergravity theory in the Jordan frame. Poincar\'{e} supergravity is recovered by choosing the K\"{a}hler potential to be $K=-\ln(-N/3)$.
\begin{align}
K&=-3\alpha\ln\left( 1-\frac{|S|^2+|\Phi|^2}{3}+\frac{g|S|^4}{3-|\Phi|^2} \right)\\W&=Sf\left( \frac{\Phi}{\sqrt{3}} \right)\left( 3-\Phi^2 \right)^{(3\alpha -1)/2}
\end{align} 
 If one looks back at the F-term potential in Eq. \hyperref[eq:F_pot]{5}, there are two features that would usually be troublesome. The $e^{K}$ is responsible for the $\eta$-problem in supergravity inflation since large values of the inflaton field would make the potential too steep to allow slow-roll inflation (especially in models with a minimal K\"{a}hler with $K\sim \Phi\bar{\Phi}$). The other troublesome feature is the $-3|W|^2$ term, which makes the potential negative-definite also for large values of $\phi$. One particular way of evading the $\eta$-problem would be to construct a $K(\Phi,\bar{\Phi})$ which is shift symmetric in the chiral fields \cite{yamaguchi,ketov2014}. The choice of the logarithmic K\"{a}hler in the $\alpha$-attractor model ensures that this does not become an issue since the prefactor simplifies to a polynomial term with a negative power. Now, to realize inflation,  $\Re\Phi=\phi$ is identified as the inflaton and the trajectory is stabilized along $\Im\Phi=S=0$.\footnote{Stabilizing the trajectory along $S=0$ also ensures that the negative-definite term does not create any problems.} Then the effective Lagrangian reduces to
\begin{equation}
\frac{1}{\sqrt{-g}}\mathcal{L}= \frac{1}{2}R-\frac{\alpha}{(1-\phi^2/3)^2}(\partial\phi)^2 - f^2\left( \frac{\phi}{\sqrt{3}} \right) 
\end{equation}
As stated previously, the scalar potential reduces to the square of the arbitrary holonomic function. Some modifications, however, need to be made since the kinetic term is non-canonical.\footnote{As stated in \cite{yamaguchi}, one of the ways of implementing inflation in supergravity is through the use of non-canonical kinetic terms.} The field redefinition $\frac{\phi}{\sqrt{3}}=\tanh\frac{\varphi}{\sqrt{6\alpha}}$ is used to canonically normalize the Lagrangian. Then, the scalar potential is simply expressed as
\begin{equation}
V(\varphi)=f^{2}\left( \tanh\frac{\varphi}{\sqrt{6\alpha}} \right)
\end{equation}
PBH production using $\alpha$-attractors has been previously studied in \cite{dalianis} where the form of $f(\phi)$ used were polynomial and sinusoidally modulated in nature. In much the same way, the deformed Starobinsky potential, as previously mentioned, can be incorporated into the $\alpha$-attractor framework. Taking the arbitrary holonomic function to be $f(\phi)=V_{0}^{1/2}\left( 1+a_{1}-e^{-a_{2}\phi}-a_{1}e^{-a_{3}\phi^2} \right)$ and canonically normalizing the field $\phi$, the end result is
\begin{equation}
V(\varphi)=V_{0}\left[ 1+a_{1}-\exp\left( -a_{2}\tanh\frac{\varphi}{\sqrt{6\alpha}} \right)-a_{1}\exp\left(-a_{3}  \tanh^2\frac{\varphi}{\sqrt{6\alpha}}\right) \right]^2
\end{equation}
The presence of the hyperbolic tangent in the potential ensures that it becomes flat for asymptotically large values of $\varphi$. Large field inflation $\varphi\gg 1$ corresponds to $\phi\approx\sqrt{3}$. In particular, the relation between $\phi$ and $\varphi$ is completely analogous to that between $v$ and the rapidity in special relativity \cite{kallosh3}. Moreover, this redefined potential inherits the plateau-like region from its previous incarnation and is thus suitable for the study of PBH formation. The $\alpha$-attractor models, at large $N$, predict the following for the scalar spectral index ($n_{s}$) and the scalar-to-tensor ratio ($r$)
\begin{equation}\label{eq:ns_alpha}
n_{s}\approx1-\frac{2}{N}\;\;\;\;\;\;\;r\approx\alpha\frac{12}{N^2}
\end{equation}
with $N$ representing the number of e-folds of observable inflation. These expressions predict values for $(n_{s},r)$ which fit observational data very well for $N\sim 56$. However, we should keep in mind that the presence of the USR phase will likely change these predictions.
\subsection{Comments on fine-tuning}
\begin{figure}
\centering
\includegraphics[scale=0.7]{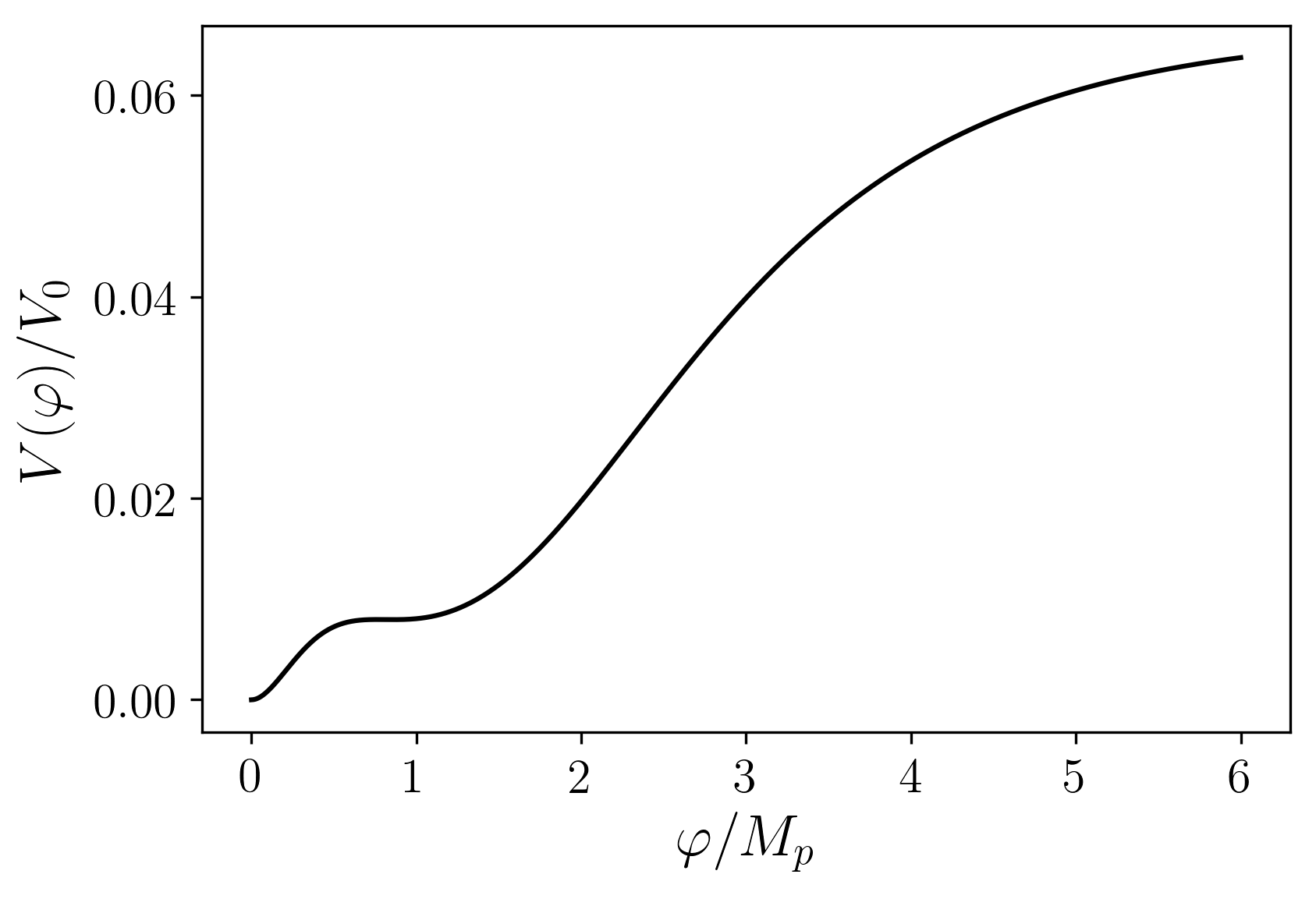}
\caption{Form of the potential for parameter set no. 2.}
\label{fig:figure2}
\end{figure}

PBH formation investigated using inflation models with an ultra slow-roll region usually require a lot of fine-tuning. The presence of a plateau-like region in the inflaton potential is not sufficient to guarantee a large enhancement in the curvature power spectrum. The inflaton may simply get stuck in that region, allowing inflation to go on indefinitely. Moreover, the inflaton may traverse the USR region but only amplify the power spectrum by a small amount. Thus, the parameters should be chosen such that the desired criteria are met. In this model, parameters simply satisfying the constraints in Eq. \hyperref[eq:constraints]{2} run into the same set of problems. Notable among them are sets of parameters that produced a longer period of USR, where the inflaton did not slow down enough and the resulting power spectra at the required scales were increased only up to $\mathcal{P}_{\mathcal{R}}\sim 10^{-5}$. A power spectrum peaking at such a value may be sufficient for PBH formation in an early matter dominated era, which is not under consideration.\\
\indent However, as stated in \cite{bellido3,ballesteros}, the problem can be resolved if the flat plateau is deformed into one with a local minimum. The inflaton slows down near the local minimum before overshooting the local maximum and large curvature perturbations are generated. The inflaton then rolls down towards the true minimum at $\varphi=0$, ending inflation. This can be achieved by modifying the second constraint equation as follows
\begin{equation}
a_{3}\longrightarrow a_{3}=\left( 1-\gamma \right)\frac{a_{2}\tilde{\varphi}+1}{2\tilde{\varphi}^2}
\end{equation}
The factor $\gamma$ is a small number and might depend on the choice of $\tilde{\varphi}$. It should be noted that the inflaton dynamics also becomes extremely sensitive to variations in the values chosen for $\gamma$, not unlike other parameters in the model. For this work, two parameter sets have been considered, which are summarized in table \hyperref[Tab:table1]{I} and the potential plotted for parameter set 2 in Fig. \hyperref[fig:figure2]{2}:
\begin{table}[h]
\centering
\begin{tabular}{|l|l|l|l|l|l|}
\hline
\# & $\alpha$ & $a_{1}$    & $a_{2}$      & $a_{3}$   & $V_{0}$             \\ \hline
1  & 1        & -0.3008933 & $\sqrt{2/3}$ & 6.1548514 & $4.7\times 10^{-9}$ \\ \hline
2  & 1        & -0.3009    & $\sqrt{2/3}$ & 6.1548416 & $4.7\times 10^{-9}$ \\ \hline
\end{tabular}
\caption{A list of two parameter sets for $V(\varphi)$ that have been considered. For both sets, $\gamma=10^{-5}$ has been used. The value for $V_{0}$ should be set by imposing the CMB normalization at observable scales.}
\label{Tab:table1}
\end{table}
\section{Inflaton dynamics and curvature perturbations}
Let us consider the dynamics of a single scalar field $\varphi$ in the presence of a potential $V(\varphi)$. Minimization of the inflaton action with respect to $\varphi$ produces the equation of motion for the inflaton in the form of the Klein-Gordon equation in curved space-time. Assuming that $\varphi$ is only a function of time, the equation of motion is given by
\begin{equation}\label{eq:inflaton_dynamics}
\ddot{\varphi}+3H\dot{\varphi}+\frac{\partial V}{\partial\varphi}=0
\end{equation} 
where $H=\dot{a}/a$ is the Hubble parameter and $t$ is the cosmic time. However, it is useful to consider the evolution of $\varphi$ in terms of e-fold time where $dN=Hdt$. Substituting $dt$ with $dN$, one finds
\begin{equation}\label{eq:inflaton}
\frac{d^2\varphi}{dN^2}+3\frac{d\varphi}{dN}-\frac{1}{2}\left( \frac{d\varphi}{dN} \right)^3+\left[ 3-\frac{1}{2}\left( \frac{d\varphi}{dN} \right)^2 \right]\frac{\partial\ln V}{\partial\varphi}=0
\end{equation}
Equation \hyperref[eq:inflaton]{18} describes the evolution of the inflaton field without resorting to any of the slow-roll approximations. The differential equation can be solved numerically for any smoothly varying potential. In order to determine the end of inflation, the first Hubble flow parameter is defined
\begin{equation}
\epsilon=\frac{1}{2}\frac{\dot{\varphi}^2}{H^2}=\frac{1}{2}\left( \frac{d\varphi}{dN} \right)^2
\end{equation}
For any initial value $\varphi_{i}$, Eq. \hyperref[eq:inflaton]{18} is solved until an $N_{f}$ when $\epsilon=1$, marking the end of inflation. Two other useful quantities are the second and third Hubble flow parameters \cite{ballesteros2015large}
\begin{align}
\eta&=-\frac{\ddot{\varphi}}{H\dot{\varphi}}=\epsilon-\frac{1}{2}\frac{d\ln\epsilon}{dN}\nonumber\\
\xi&=\frac{\dddot{H}}{2H^{2}\dot{H}}-2\eta^{2}=\epsilon\eta-\frac{d\eta}{dN}
\end{align}
which help determine the scalar spectral index and its running. Slow roll inflation requires that $\epsilon\ll 1$ and $\eta \ll 1$, only reaching unity at the end of inflation. However, it will be shown that during the USR phase, $\eta$ breaks the slow-roll condition temporarily. Under the slow-roll approximation, the power spectrum of curvature perturbations can be expressed in terms of the first Hubble flow parameter \cite{Stewart:1993bc}
\begin{equation}\label{eq:pspec_sr}
\mathcal{P}_{\mathcal{R}}\simeq\frac{H^2}{8\pi^2\epsilon}
\end{equation}
Since $\epsilon$ is a measure of the velocity of the inflaton field, it might be tempting to think that Eq. \hyperref[eq:pspec_sr]{21} indeed serves as an accurate determinant of the curvature power spectrum. This is not so, since the slow-roll approximation also includes the condition that $H$ does not change substantially during inflation. Because there is an $\mathcal{O}(1)$ change in the Hubble parameter, the slow-roll approximation is not valid and the Mukhanov-Sasaki equation is solved numerically to obtain the exact power spectrum. Nevertheless, the slow-roll approximation can be effectively used to obtain important observables at CMB scales \cite{liddle2000cosmological}
\begin{align}
n_{s}&\simeq1-4\epsilon_{\star} + 2\eta_{\star}\\
\alpha_{s}&\simeq-2\xi_{\star}+10\epsilon_{\star}\eta_{\star}-8\epsilon_{\star}^2 \\
\mathcal{P}_{\mathcal{R}}&\simeq\frac{H_{\star}^{2}}{8\pi^{2}\epsilon_{\star}}\\
r&\simeq16\epsilon_{\star}
\end{align}
Here $n_{s}$, $\alpha_{s}$ and $r$ are the scalar spectral index, its running and the scalar-to-tensor ratio respectively. The stars imply that these observables have been evaluated at the CMB scale ($k=\unit[0.05]{Mpc^{-1}}$). The values of these observables will be discussed in Sec. \hyperref[sec:Press_Schechter]{IV.B}.
\subsection{The Ultra Slow-Roll period}
\begin{figure}
     \centering
     \includegraphics[scale=0.55]{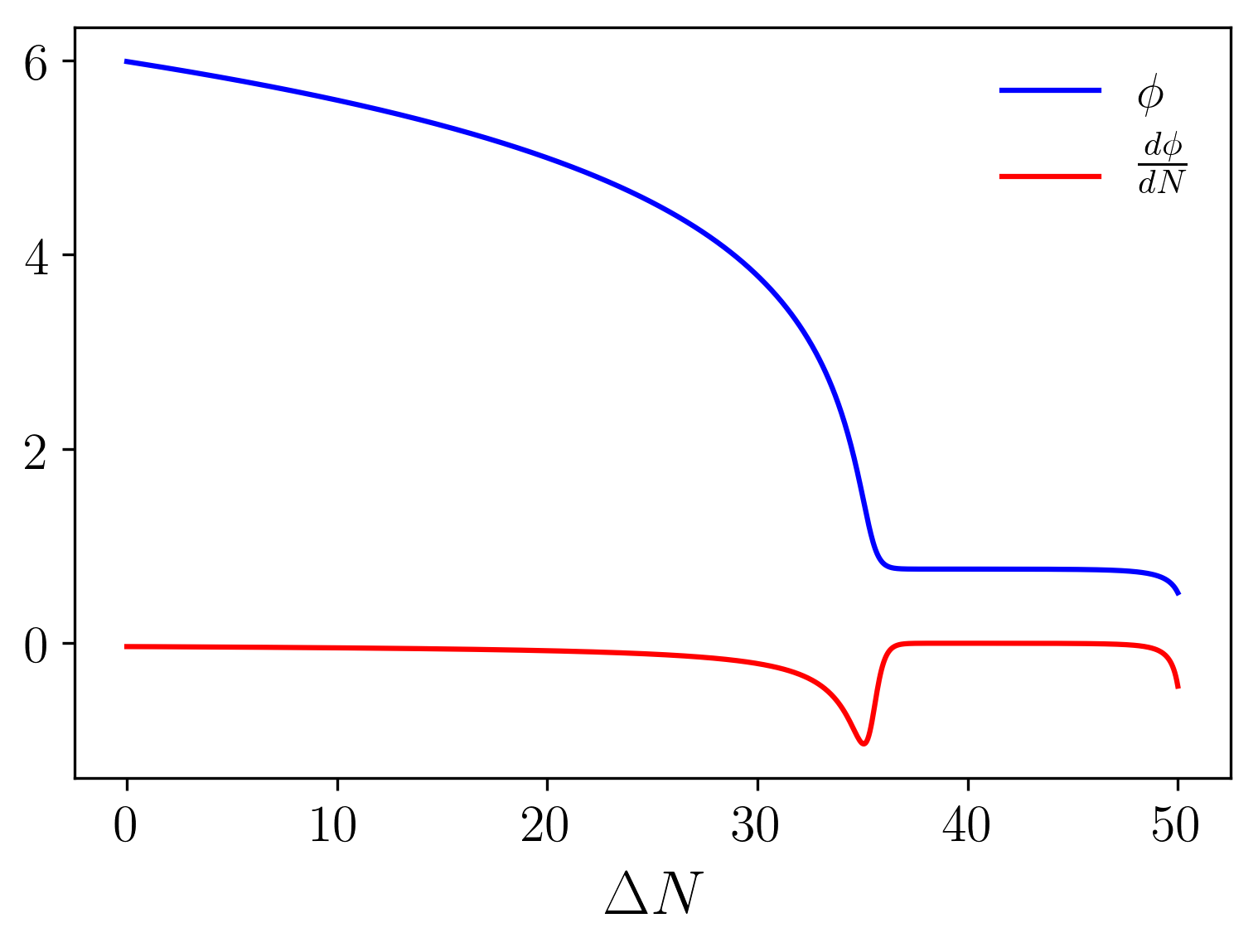}
     \includegraphics[scale=0.55]{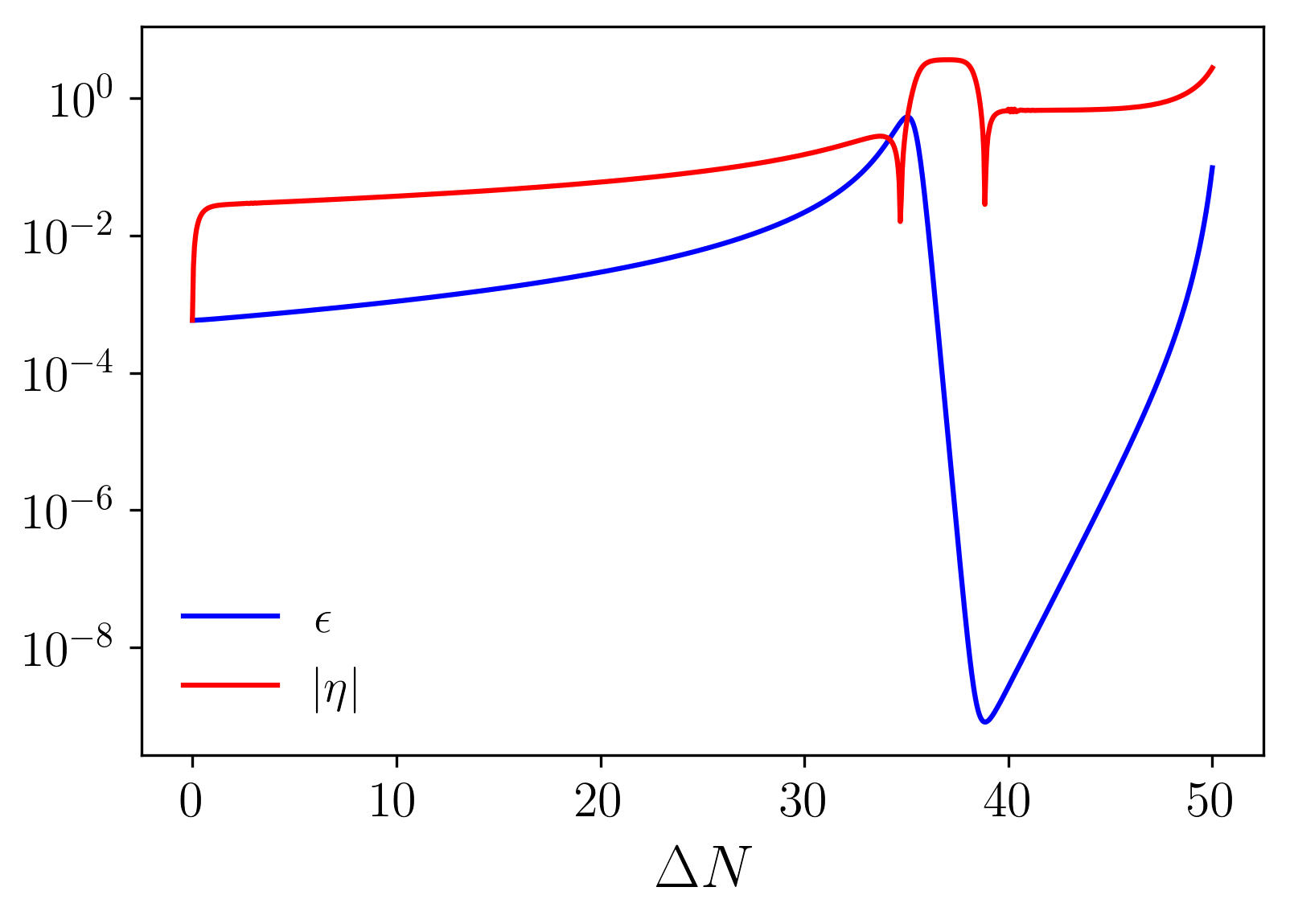}
        \caption{\textit{Left}: The background evolution of $\varphi$ (blue) and $\frac{d\varphi}{dN}$ (red). \textit{Right}: Variation of the first two Hubble flow parameters - $\epsilon$ (blue) and $|\eta|$ (red). The evolutions have been calculated with respect to amount of observable inflation $\Delta N$. The effect of the USR period is evident in these graphs. We also see that $\eta >1$ during this period.}
        \label{fig:figure3}
\end{figure}

For a substantial period, from the time the CMB scales became superhorizon, the slow-roll approximation works rather well. The Hubble parameter remains approximately constant and the power spectrum displays the nearly scale-invariant characteristics that we know of. This slow-roll evolution continues for a period after which the evolution of $\varphi$ enters the USR period. The location and duration of USR is model dependent but mostly it lasts for a short period of time near the end of inflation.\\
\indent In the USR phase, the inflaton arrives at the plateau around the inflection point where $\partial_{\varphi}V\approx 0$ and the term with the derivative of the potential in Eq. \hyperref[eq:inflaton_dynamics]{17} drops out. Also, the acceleration term in the Klein-Gordon equation can no longer be ignored \cite{dimo}. This results in
\begin{equation}
\ddot{\varphi}+3H\dot{\varphi}\approx 0
\end{equation}
which leads to
\begin{equation}
\eta = -\frac{\ddot{\varphi}}{H\dot{\varphi}}\approx 3
\end{equation}
Under the SR approximation, both $\epsilon$ and $\eta$ should be less that unity. Hence, $\eta$ breaks slow-roll, if only temporarily.\footnote{It is also possible that $\epsilon$ violates the slow-roll condition. This is what happens to $\epsilon$ for the polynomial potential in \cite{dalianis}.} Through Eq. \hyperref[eq:pspec_sr]{21}, one can also argue in favour of an increase of the power spectrum. During this phase, $\varphi\sim e^{-3\delta N}$ and, as a consequence, $\mathcal{P}_{\mathcal{R}}\sim e^{6\delta N}$. The Hubble flow functions also have analogues in terms of the potential and their derivatives, called the potential slow-roll parameters $\epsilon_{\text{V}}$ and $\eta_{\text{V}}$. They are expressed as
\begin{equation}
\epsilon_{\text{V}}=\frac{1}{2}\left( \frac{\partial_{\varphi}V}{V} \right)^2\;\;\;\;\;\;\;\;\eta_{\text{V}}=\frac{\partial_{\varphi\varphi}V}{V}
\end{equation}
The expectation is that $\epsilon\simeq\epsilon_{\text{V}}$ and $\eta\simeq\eta_{\text{V}}$ during slow-roll. Numerical solutions show that it is true up to the beginning of the USR period where the potential slow-roll parameters display departures from their Hubble counterparts. With these considerations, the background evolution of the inflaton is solved and the first two Hubble flow parameters are computed. The results are plotted in Fig. \hyperref[fig:figure3]{3} for parameter set 2.
\begin{table}[h]
\centering
\begin{tabular}{|l|l|l|l|l|l|l|l|}
\hline
\# & $\alpha$ & $a_{1}$    & $a_{2}$      & $a_{3}$   & $V_{0}$             & $\varphi_{\star}$ & $\Delta N$ \\ \hline
1  & 1        & -0.3008933 & $\sqrt{2/3}$ & 6.1548514 & $4.7\times 10^{-9}$ & 5.896             & 50.3       \\ \hline
2  & 1        & -0.3009    & $\sqrt{2/3}$ & 6.1548416 & $4.7\times 10^{-9}$ & 5.896             & 50.9       \\ \hline
\end{tabular}
\caption{The two parameter sets and the corresponding amount of observable inflation. $\varphi_{\star}$ represents the inflaton field value when CMB scales become superhorizon.}
\end{table}
\subsection{Enhancement of curvature perturbations}
\begin{figure}[t]
\centering
\includegraphics[scale=0.8]{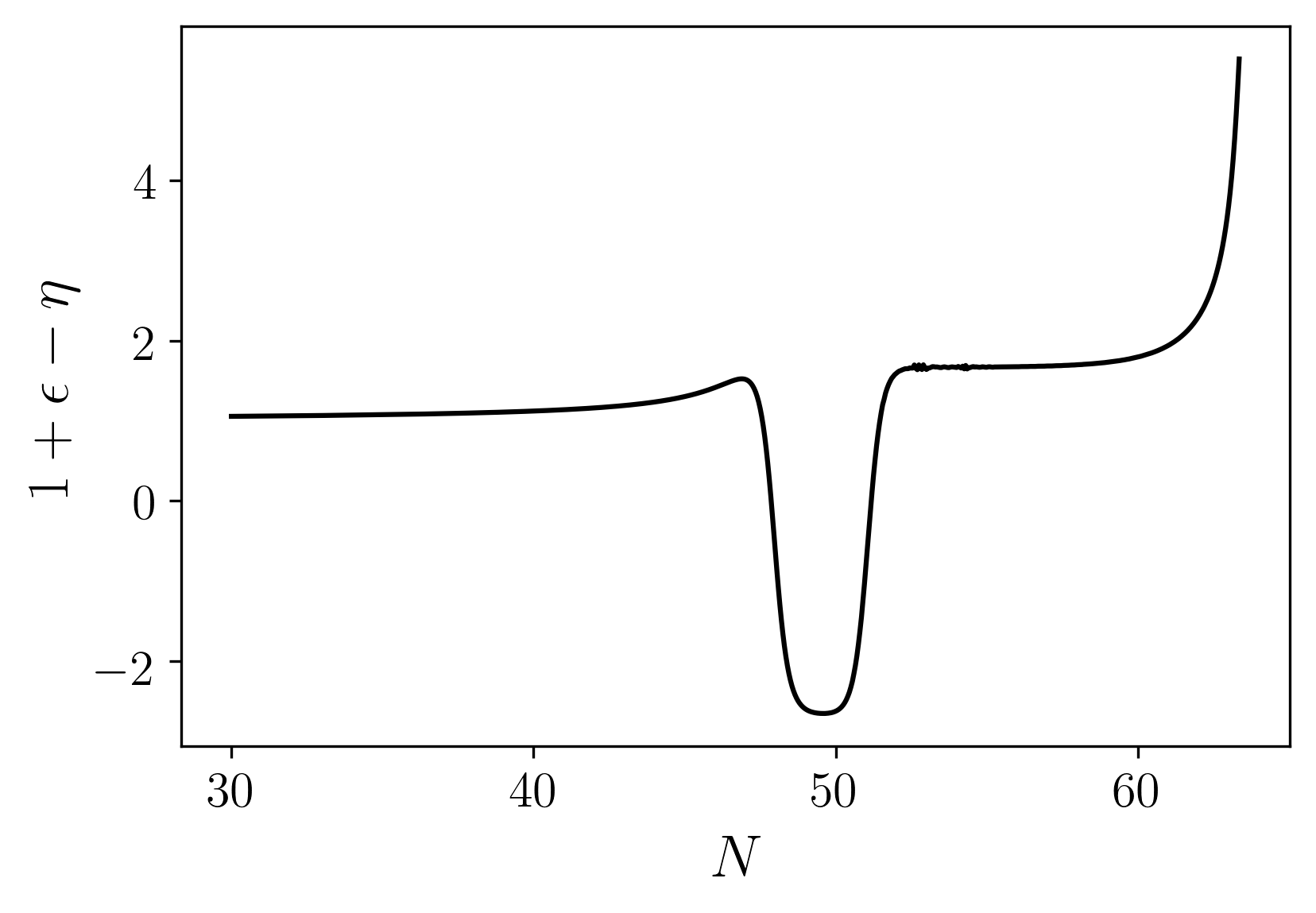}
\caption{Plot of $1+\epsilon-\eta$ (for parameter set 2) showing its contribution to the growing modes during USR.}
\label{fig:z'_over_z}
\end{figure}

During inflation, the inflaton dominates the energy density of the universe. Fluctuations of the inflaton then produce perturbations of the stress-energy tensor which, in turn, induce metric perturbations. The metric perturbations of the scalar kind are responsible for the energy density fluctuations in the post-inflationary stages. The curvature perturbation can be understood as deviations of the spatial part of the metric away from the FLRW background $g_{ij}=a({\tau})\left[ (1-2\Psi)\delta_{ij}+h_{ij} \right]$. In fact, the Ricci scalar on the spatial slices is expressed as $^{(3)}R=\frac{4}{a^2}\nabla^{2}\Psi$. In conventional notation, $\Psi$ is not a gauge-invariant quantity. The gauge-invariant curvature perturbation,\footnote{In literature, another commonly used gauge-invariant curvature perturbation is $\zeta$, which coincides with $\Psi$ on constant energy hypersurfaces $\delta\rho=0$. Outside the horizon, $\mathcal{R}=\zeta$ and both are conserved.} $\mathcal{R}$, coincides with $\Psi$ in the comoving gauge where $\delta\varphi=0$. The evolution of the curvature perturbation can be derived using the ADM formalism \cite{khachru}.\\
\indent In Fourier space, the evolution of the curvature perturbations is dictated by the Mukhanov-Sasaki (MS) equation \cite{leach2001,leach2}
\begin{equation}
\mathcal{R}''_{k}+2\frac{z'}{z}\mathcal{R}'_{k}+k^2\mathcal{R}_{k} =0
\end{equation} 
where the primes denote derivatives with respect to the conformal time, $z=a\varphi'/\mathcal{H}$ and $z'/z=2aH\left( 1+\epsilon-\eta \right)$. The MS equation appears to take the form of a damped harmonic oscillator and has two well-defined regions characterised by the comoving Hubble horizon $(aH)^{-1}$. The second term in the equation can be thought of as a friction term which drops out during subhorizon evolution $(k\gg aH)$, describing modes with oscillatory features. At the other end, over superhorizon scales, $k\ll aH$, the solution assumes the form
\begin{equation}\label{eq:MS_eq0}
\mathcal{R}_{k\ll aH}=C_{2}+C_{1}\int d\tau\;e^{-\int d\tau'2aH(1+\epsilon-\eta)}
\end{equation}
where $C_{1}$ and $C_{2}$ are constants of integration. Equation \hyperref[eq:MS_eq0]{30} can be simplified using the fact that $z'/z=(\ln z)'=aH(1+\epsilon-\eta)$. Then,
\begin{equation}
\mathcal{R}_{k\ll aH}=C_{2}+C_{1}\int\frac{d\tau}{z^2}
\end{equation}
We are now in a position to understand, at least semi-quantitatively, how the curvature perturbation is enhanced. Apart from a constant, the solution of the MS equation has a time-dependent component appearing in the integral that depends on the Hubble flow parameters. For scales that left the horizon during the SR phase, $\epsilon$ and $\eta$ are small and the term $1+\epsilon-\eta$ is positive\footnote{In fact, in the usual slow-roll, $\epsilon$ and $\eta$ are small and do not change appreciably and the second term in Eq. \hyperref[eq:MS_eq0]{30} only decays slowly, leading to a nearly scale-invariant power spectrum.}, meaning that the second term produces a decaying contribution. However, beyond the slow-roll approximation (eg. in the USR phase), $\eta$ could become large and turn the decaying mode into a growing one (Fig. \hyperref[fig:z'_over_z]{4}). If such a period exists, it can cause the curvature power spectrum to grow at the time of horizon exit. In fact, this is precisely what had been shown in the previous section where the second Hubble flow parameter becomes $\eta\approx 3$ during USR.\\
\indent We now turn our attention to the exact numerical solution of the MS equation. To that end, the equation is recast to a different form using the field redefinition $v_{k}=z\mathcal{R}_{k}$ with $z=ad\varphi/dN$. The reason for the field redefinition is simply that the evolution equation resembles that of a harmonic oscillator with a time-dependent frequency term in conformal time and the delineation between subhorizon and superhorizon evolution becomes more apparent. The field redefined Mukhanov-Sasaki equation reads
\begin{equation}\label{eq:MS_eq}
\frac{d^2v_{k}}{dN^2}+(1-\epsilon)\frac{dv_{k}}{dN}+\left[ \left( \frac{k}{aH} \right)^2 +(1+\epsilon-\eta)(\eta-2)-\frac{d}{dN}(\epsilon-\eta) \right]v_{k}=0
\end{equation}
Each Fourier mode of the curvature perturbation starts out from subhorizon scales where the evolution can be treated like that of a free field in Minkowski space and the modes can be expanded as plane waves. The modes $v_{k}$ start out in the Bunch-Davies vacuum state
\begin{equation}\label{eq:BD_vac}
v_{k}\longrightarrow \frac{e^{-ik\tau}}{\sqrt{2k}}
\end{equation}
in the asymptotic past. Then, each $v_{k}$ evolves until it becomes superhorizon, thereby reaching a constant value. Using these values, the power spectrum of curvature perturbations is defined
\begin{equation}\label{eq:PSpectrum_exact}
\mathcal{P}_{\mathcal{R}}=\frac{k^3}{2\pi^2}\left| \frac{v_{k}}{z} \right|^{2}_{k\ll aH}
\end{equation}
For a numerical solution of Eq. \hyperref[eq:MS_eq]{32}, it is expedient to separate $v_{k}$ into real and imaginary parts and solve them using the background evolution solutions. Equation \hyperref[eq:BD_vac]{33} can also be used to set initial conditions for both the real and imaginary parts of $v_{k}$ and their first derivatives.\\
\indent Depending on the number of e-folds of observable inflation, the range of $k$ is determined via
\begin{equation}
k=k_{\star}\frac{H(N_{\text{end}})}{H_{\star}}e^{\Delta N}
\end{equation}
where $k_{\star}$ is the pivot scale (when CMB scales left the horizon) and $\Delta N=N_{\text{end}}-N_{\star}$ is the amount of observable inflation. For each $k$, the evolution of $v_{k}$ is initiated deep inside the horizon or when $k=100a(N_{i})H(N_{i})$ with the following initial conditions
\begin{equation}
\text{Re}(v_{k})=\frac{1}{\sqrt{2k}},\;\;\text{Im}(v_{k})=0,\;\;\text{Re}\left( \frac{dv_{k}}{dN} \right)=0,\;\;\text{Im}\left( \frac{dv_{k}}{dN} \right)=-\sqrt{\frac{k}{2}}\frac{1}{10^{-2}k}
\end{equation} 
Once the modes exit the horizon, they become constant and are evolved until $k=0.01a(N_{f})H(N_{f})$. For each $k$, this superhorizon value is extracted and used in Eq. \hyperref[eq:PSpectrum_exact]{34} to determine the power spectrum of curvature perturbations. The choice of duration of sub to superhorizon evolution used here is rather arbitrary (but not an uncommon one) and it is usually suggested that the modes are allowed at least 5 e-folds of sub and superhorizon evolution.\\
\begin{figure}[t]
     \centering

         \includegraphics[scale=0.8]{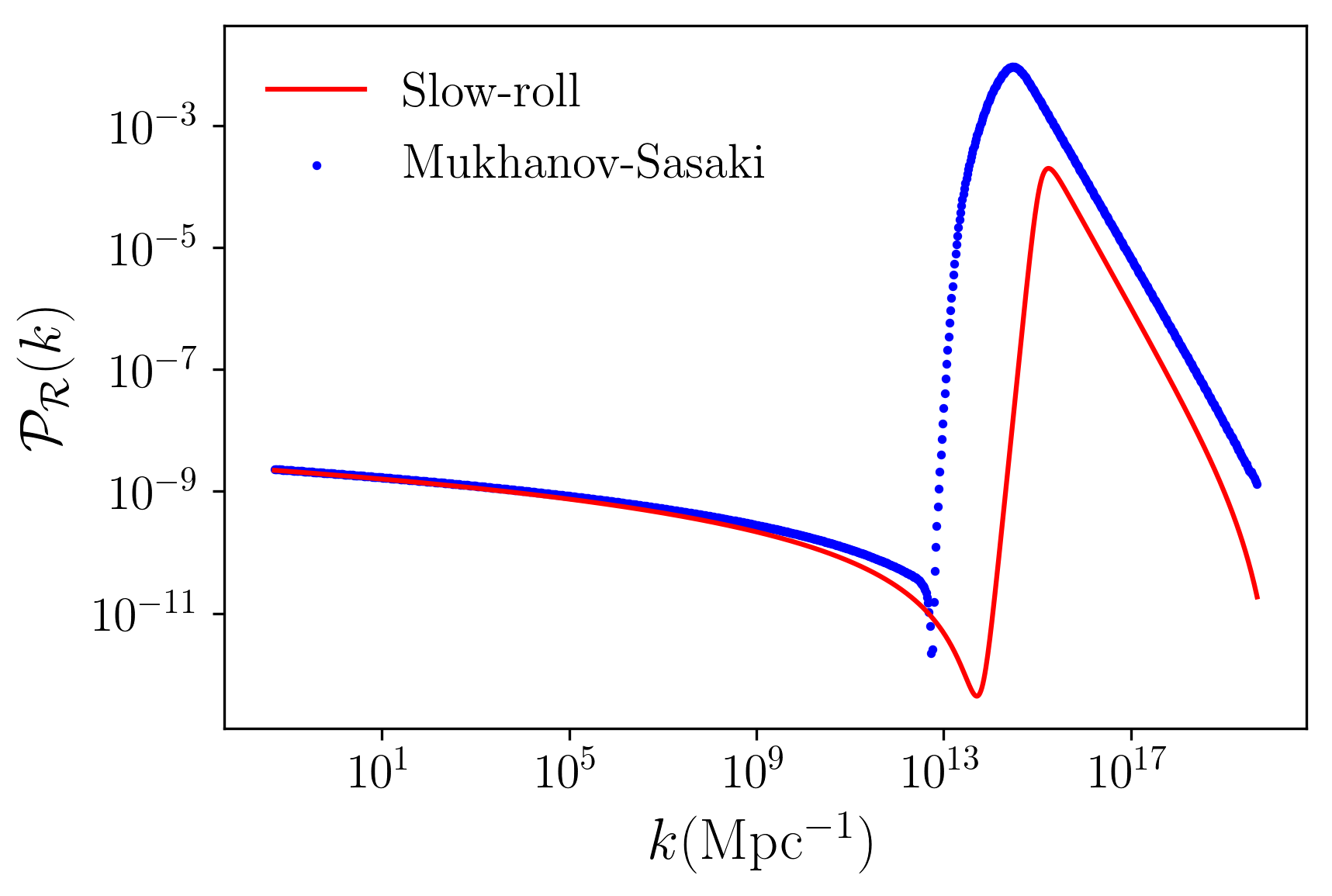}

     \vfill
          \includegraphics[scale=0.8]{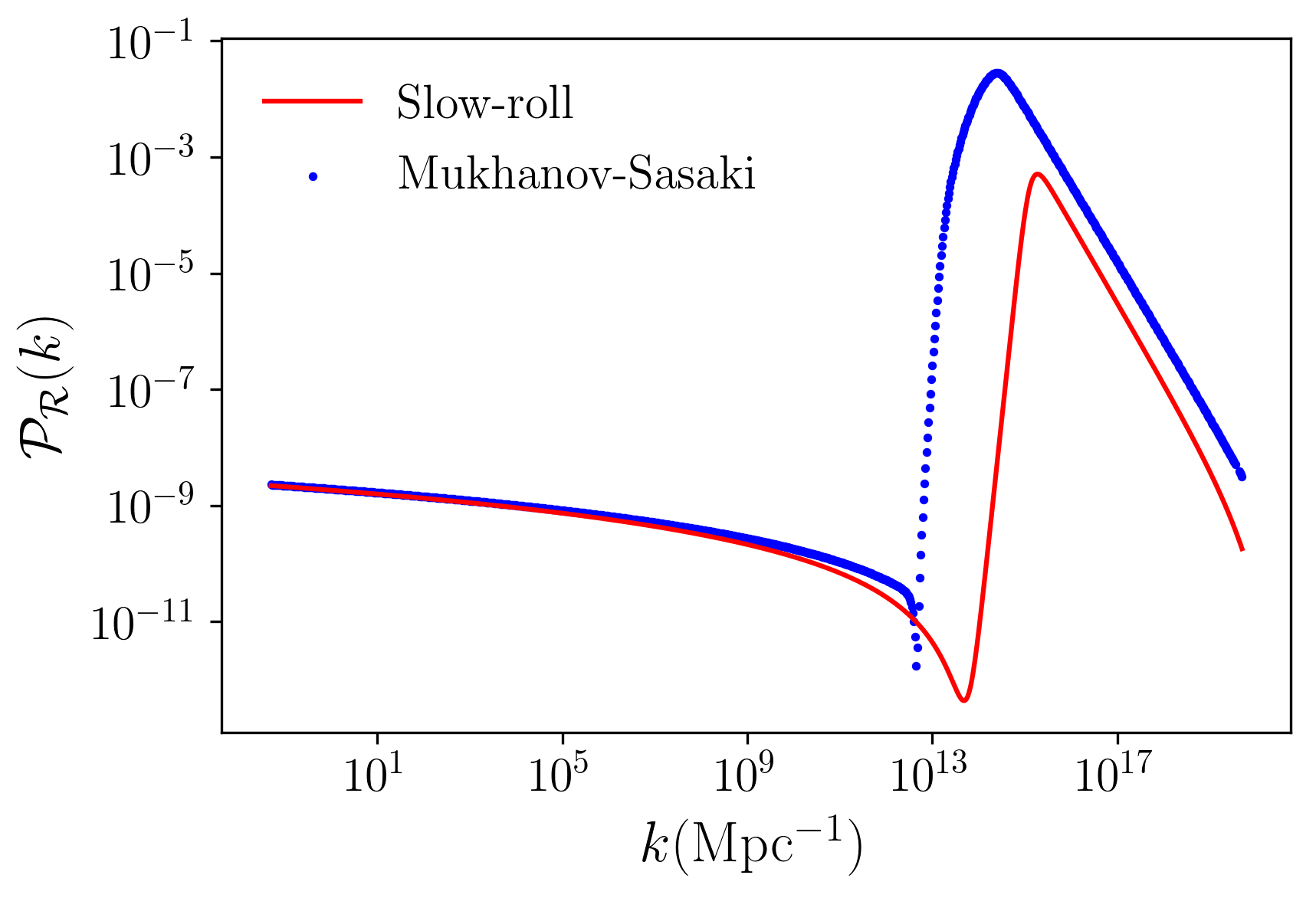}
        \caption{Plots of the curvature power spectra for parameter set 1 (top panel) and set 2 (bottom panel). The red solid lines represent the slow-roll approximation while the blue dotted ones represents the exact $\mathcal{P}_{\mathcal{R}}$ using the Mukhanov-Sasaki equation.}
        \label{fig:figure4}
\end{figure}
\indent The power spectra for curvature perturbations have been plotted in Fig. \hyperref[fig:figure4]{5} for both parameter sets. The two curves agree very well for large scales when the inflaton dynamics is in slow-roll. However, it is clear from the graphs that, once USR takes over, the slow-roll approximation for $\mathcal{P}_{\mathcal{R}}$ underestimates the numerical calculation by at least one order of magnitude. The peaks in $\mathcal{P}_{\mathcal{R}}$, too, occur at higher values of $k$ under the approximation. Although PBH formation can be examined using the approximated power spectra, we shall see that such a scenario implies the collapse of smaller overdensities. The placement of the peaks towards higher $k$ would also imply the formation of PBHs with $M<\unit[10^{16}]{g}$. Such PBHs will have already evaporated or are at their final stages.
\section{PBH formation}
In this section, the formation of PBHs, when high curvature perturbation modes re-enter the horizon after inflation, will be studied. The dependence of the PBH mass on $k$ will be derived followed by the formation fraction via the Press-Schechter formalism. The amount of the PBH energy density that goes into dark matter will also be discussed using the two parameter sets.
\subsection{PBH mass}
During inflation, the comoving Hubble horizon $(aH)^{-1}$ shrinks and all the scales $k^{-1}$ leave the horizon at some point. When the inflationary expansion stops, the comoving Hubble horizon starts growing again and these scales start re-entering the horizon- with the smallest scales re-entering first. The curvature perturbations related to scales $k^{-1}$ influence the density fluctuations of the radiation plasma and, if they are large enough, can cause gravitational collapse forming PBHs. Now we attempt to find a relationship between the mass of PBHs and the collapse related to the re-entry of scales $k^{-1}$. This is done with the assumption that the mass of a PBH via the collapse of a curvature perturbation mode is proportional to the horizon mass
\begin{equation}\label{eq:pbh_mass}
M(k)=\frac{4\pi}{3}\gamma\rho H_{k}^{-3}\bigg\rvert_{k=aH}=\gamma M_{\text{eq}}\frac{g_{\star}(T_{k})}{g_{\star}(T_{\text{eq}})}\left( \frac{T_{k}}{T_{\text{eq}}} \right)^{4}\left( \frac{H_{\text{eq}}}{H_{k}} \right)^3
\end{equation}
where $\rho$ is the energy density of the universe during the collapse and $\gamma$ is the efficiency factor, usually set to 0.2. The second equality has been obtained by first taking a ratio of $M(k)$ with the horizon mass at matter-radiation equality $M_{\text{eq}}$ and using the fact that, during the radiation epoch, $\rho\propto g_{\star}(T)T^{4}$. Also, assuming entropy conservation $g_{s}(T)T^{3}a^{3}=\text{constant}$, we arrive at
\begin{equation}
M(k)\simeq1.6\times10^{18}\;\text{g}\;\left( \frac{\gamma}{0.2} \right)\left( \frac{g_{\star}(T_{k})}{106.75} \right)^{-1/6}\left( \frac{k}{5.5\times 10^{13}\;\text{Mpc}^{-1}} \right)^{-2}
\end{equation}
In the above derivation, it has been assumed that the effective number of adiabatic and entropic degrees of freedom are equal ($g_{\star}=g_{s}$). The fact that the horizon mass at matter-radiation equality is $M_{\text{eq}}\simeq 5.6\times 10^{50}\;\text{g}$, during which time $k_{\text{eq}}=0.07\Omega_{\text{m}}h^{2}\;\text{Mpc}^{-1}$, has also been used. Moreoever, $g_{\star}(T_{\text{rad}})=106.75$ and $g_{\star}(T_{\text{eq}})=3.36$.
\subsection{Press-Schechter formalism and PBH formation fraction}
\label{sec:Press_Schechter}
\begin{figure}
     \centering
\includegraphics[scale=1]{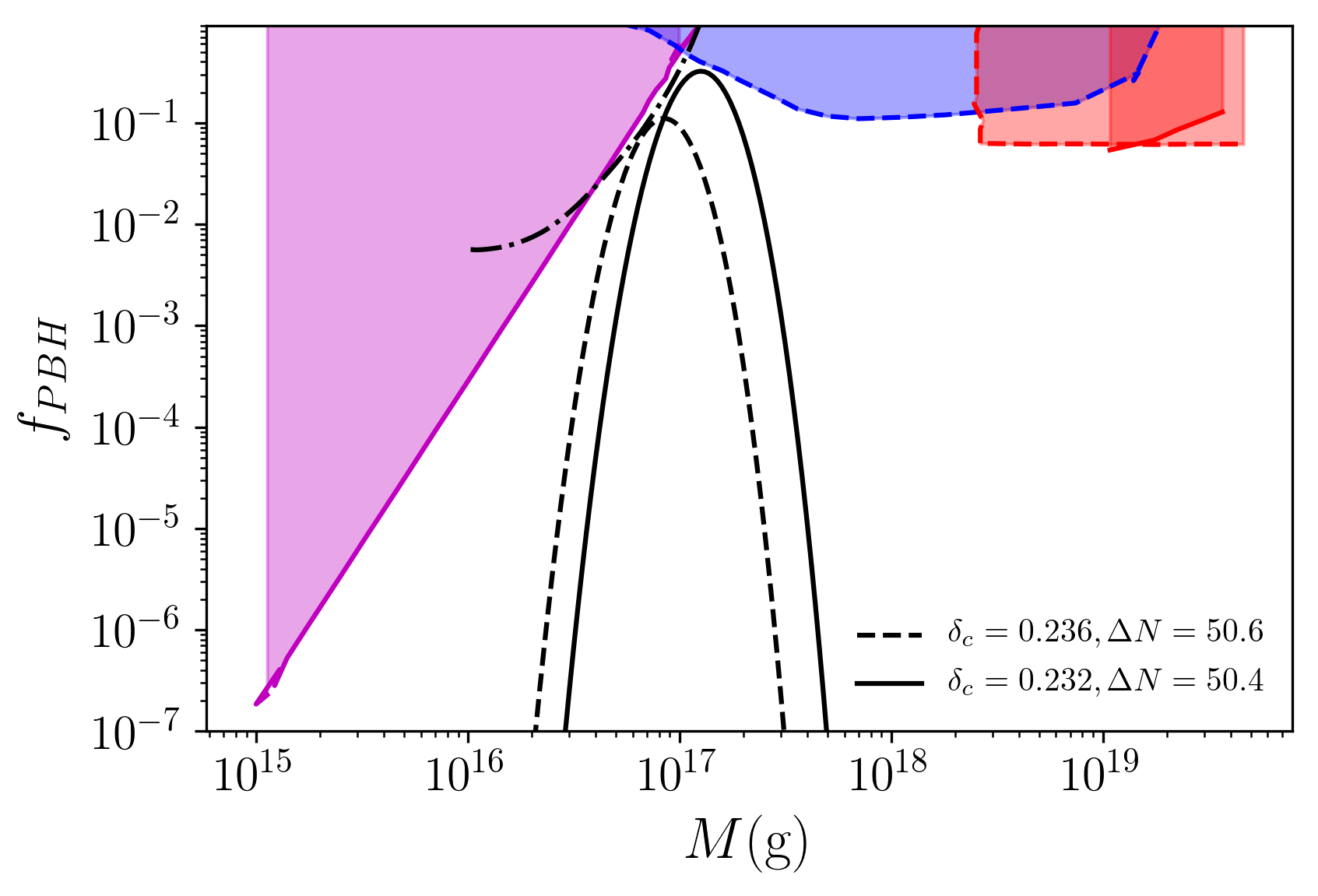}
\includegraphics[scale=1]{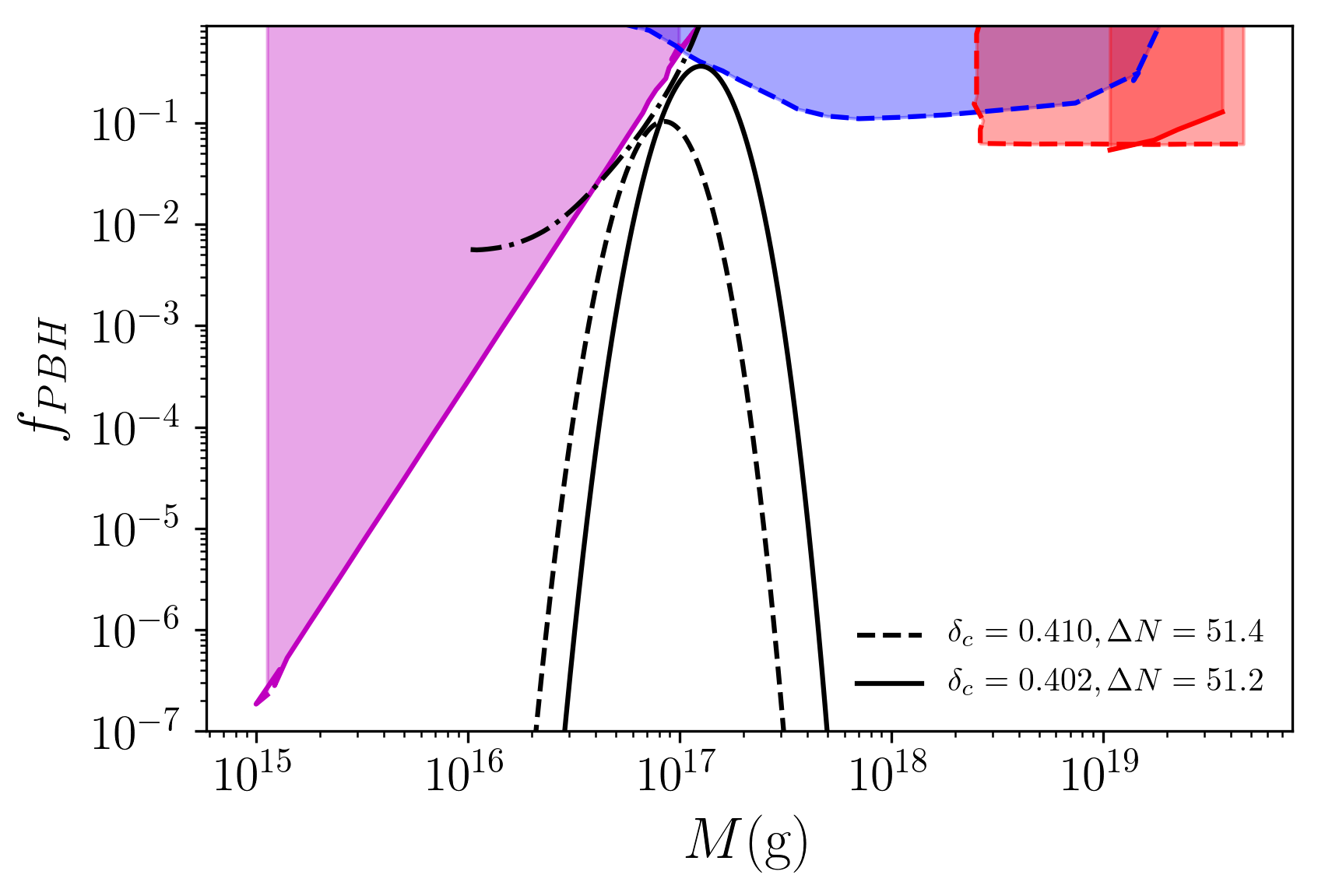}
        \caption{The $f_{\text{PBH}}$ for parameter set 1 (\textit{top}) and 2 (\textit{bottom}) for two different $\Delta N$. The shaded portions represent the different constraints that are relevant in that particular mass range: extragalactic $\gamma$-ray background \cite{carr2016constraints} (magenta), $e^{\pm}$ annihilation and Galactic Center $\unit[511]{keV}$ gamma-ray line (black dashed-dotted) \cite{Laha:2019ssq}, femtolensing of $\gamma$-ray bursts \cite{katz2018femtolensing} (blue) and white dwarf and neutron star capture (red). The horizontal axis has been cut-off to contain only the relevant mass range. Details on the constraints are discussed in Sec. \hyperref[sec:section4D]{IV.D}.}
        \label{fig:figure6}
\end{figure}

Assuming PBH formation takes place in the radiation epoch, one can consider overdense regions that are able to overcome radiation pressure and gravitationally collapse. Denoting the overdensities as $\delta\equiv\delta\rho/\rho$, they should be over a certain threshold, $\delta_{c}$, for PBH formation. For curvature perturbations obeying Gaussian statistics, the Press-Schechter formalism \cite{press1974} provides a method for computing the fraction of PBHs formed with mass $M(k)$ over the total radiation energy density.
\begin{align}\label{eq:form_frac}
\beta(M(k))\equiv\frac{\rho_{\text{PBH}}}{\rho_{\text{rad}}}&=\int_{\delta_{c}}^{\infty}\frac{d\delta}{\sqrt{2\pi}\sigma(M(k))}\exp\left( -\frac{\delta^2}{2\sigma^{2}(M(k))} \right)\nonumber\\
&=\frac{1}{2}\text{erfc}\left( \frac{\delta_{c}}{\sqrt{2}\sigma(M(k))} \right)
\end{align}
Density perturbations with $\delta >\delta_{c}$ are the ones that are able to overcome radiation pressure and collapse gravitationally. The integral should be ideally carried out between $\delta_{c}$ and some $\delta_{\text{max}}$. However, since the formation fraction is exponentially sensitive to the threshold density, integration up to infinity is a very good approximation.\\
\indent One of the earliest analytical estimates of the threshold was derived by Carr \cite{carr1} using simple Jeans instability analysis where he found $\delta_{c}\approx 1/3$ for PBH formation during the radiation era. Over the years, with more sophisticated numerical simulations, different values of $\delta_{c}$ have been calculated \cite{harada,shibata1999,musco2009,germani2019,niemeyer}. While studying PBH formation using inflation models, the characteristics of the curvature power spectrum play an important role in determining the threshold. It is, at this point, that the concept of an enhanced curvature power spectrum becomes important as it influences $\delta_{c}$.\footnote{PBH formation during a period of matter domination can be easier to accomplish with smaller enhancements in the curvature power spectrum (since $w=0$). A scenario like this can be realized in models with a non-instantaneous reheating phase.} Now, the density perturbations $\delta$ depend on the curvature perturbations through the following equation (up to linear order) \cite{liddle2000cosmological}
\begin{equation}
\delta(\bm{x},t)=\frac{2(1+w)}{5+3w}\frac{1}{(aH)^2}\nabla^{2}\mathcal{R}(\bm{x},t)
\end{equation}
or in Fourier space
\begin{equation}
\delta_{k}(t)=-\frac{2(1+w)}{5+3w}\left( \frac{k}{aH} \right)^{2}\mathcal{R}_{k}(t)
\end{equation}
Here $w$ is the equation of state parameter of the epoch of re-entry and $w=1/3$ during radiation era. We see that these density perturbations are initially small when the scales under consideration are superhorizon due to the $(k/aH)^2$ dependence. As the comoving Hubble horizon starts to grow, $\delta$ starts to increase and the threshold is reached a short time after $k^{-1}$ enters the horizon. Now, we see in Eq. \hyperref[eq:form_frac]{39} that $\beta(M)$ depends on the mass variance $\sigma(M)$, which is obtained by first coarse-graining over a scale $q^{-1}$ with a suitably chosen window function. There are several choices for this window function and a commonly employed one is the Gaussian filter. In Fourier space, a Gaussian filter looks like
\begin{equation}
W\left(\frac{q}{k}\right)=e^{-q^{2}/2k^{2}}
\end{equation}
Then, the coarse-grained (smoothed) mass variance is defined as
\begin{equation}
\sigma^{2}(M(k))=\frac{16}{81}\int\frac{dq}{q}W^{2}\left( \frac{q}{k} \right)\left( \frac{q}{k} \right)^{4}\mathcal{P}_{\mathcal{R}}(q)
\end{equation}
Using these, the PBH formation fraction can be calculated for the curvature perturbation modes that become subhorizon during the radiation epoch. After formation, the energy density of PBHs increase linearly with the scale factor since $\rho_{\text{PBH}}\propto a^{-3}$ and $\rho_{\text{rad}}\propto a^{-4}$. This continues until matter-radiation equality when the background energy density starts scaling like that of PBHs. Since the PBH energy densitiy stops growing from that point, it serves as a benchmark to computing the fraction of PBHs with mass $M$ over the dark matter relic density, defined as $f_{\text{PBH}}(M)$.
\begin{align}
f_{\text{PBH}}&\equiv\frac{\Omega_{\text{PBH}}(M(k))}{\Omega_{\text{DM}}}\nonumber\\
&\simeq\frac{\Omega_{\text{m}}}{\Omega_{\text{DM}}}\frac{\rho_{\text{PBH}}}{\rho_{\text{rad}}}\bigg\lvert_{\text{eq}}
\end{align}
$\Omega_{\text{m}}$ and $\Omega_{\text{DM}}$ are the total matter and dark matter relic densities. When curvature perturbations associated with modes $k$ collapse, it is assumed that a fraction $\gamma\beta(M(k))\rho$ of the energy density of the universe goes into the production of PBHs. After formation in the radiation epoch, $\beta$ grows linearly with $a$ until matter-radiation equality. Hence
\begin{align}\label{eq:fpbh}
f_{\text{PBH}}(M(k))&=\left( \frac{T_{k}}{T_{\text{eq}}}\frac{\Omega_{\text{m}}}{\Omega_{\text{DM}}} \right)\gamma\beta(M(k))\nonumber \\
&\simeq \left( \frac{\beta(M(k))}{10^{-15}} \right)\left( \frac{\gamma}{0.2} \right)^{3/2}\left( \frac{g_{\star}(T_{k})}{106.75} \right)^{-1/4}\left( \frac{M(k)}{\unit[1.6\times 10^{18}]{g}} \right)
\end{align}  
In Eqs. \hyperref[eq:pbh_mass]{37} and \hyperref[eq:fpbh]{45}, $T_{k}$ refers to the temperature of the radiation plasma at the time the scales $k^{-1}$ re-entered the horizon, collapsed and formed PBHs with mass $M(k)$. In order to compare the total energy density in PBHs in the current epoch to that of the dark matter relic density, $f_{\text{PBH}}(M)$ is integrated over the masses
\begin{equation}
f_{\text{PBH}}=\int d\ln M\frac{\Omega_{\text{PBH}}(M)}{\Omega_{\text{DM}}}
\end{equation}
For the two parameter sets, the Mukhanov-Sasaki equation is solved by allowing inflation to continue for two slightly different e-foldings of observable inflation. The two different $\Delta N$ correspond to two different $k_{\text{peak}}$ and, consequently, two different $f_{\text{PBH}}(M)$ curves for each parameter set. The results have been summarized in table \hyperref[Tab:table3]{III}, with the observables for both parameter set 1 and 2. For 1, because of the lower $\mathcal{P}_{\mathcal{R}}^{\text{peak}}$, a lower $\delta_{c}$ is required to reach the upper bound.\footnote{Since the formation fraction $\beta$ is an integral with $\delta_{c}$ as a lower limit, smaller $\delta_{c}$ means a large $\beta$. Another interpretation is that PBH formation from lower $\delta_{c}$ means collapse of overdensities which are not very rare.} Figure \hyperref[fig:figure6]{6} also shows the $f_{\text{PBH}}(M)$ for this model, resembling log-normal distributions.\\
\begin{table}[t]
\centering
\begin{tabular}{|l|l|l|l|l|l|l|l|l|}
\hline
\#                      & $\Delta N$                & $n_{s}$                      & $\alpha_{s}$                  & $r$                          & $\mathcal{P}_{\mathcal{R}}^{\text{max}}$ & $M^{\text{max}}/M_{\odot}$                 & $\delta_{c}$               & $f_{\text{PBH}}$            \\ \hline
1 & 50.6 & 0.94405 & -0.00164 & 0.00893 & 0.00929             & $4.16\times 10^{-17}$ & 0.236 & 0.0743 \\ \hline
1 & 50.4 & 0.94373 & -0.00166 & 0.00903 & 0.00930             & $6.3\times 10^{-17}$  & 0.232 & 0.2209 \\ \hline
2 & 51.4 & 0.94405 & -0.00164 & 0.00893 & 0.02789             & $4.24\times 10^{-17}$ & 0.410 & 0.0693 \\ \hline
2 & 51.2 & 0.94373 & -0.00164 & 0.00903 & 0.02804             & $6.3\times 10^{-17}$  & 0.402 & 0.2478 \\ \hline
\end{tabular}
\caption{Summary of the results obtained from the two parameter sets with two different values for $\Delta N$ as well as observables at $k=\unit[0.05]{Mpc^{-1}}$. These correspond to two different curves for $f_{\text{PBH}}(M)$ for each parameter set.}
\label{Tab:table3}
\end{table}
\indent According to their 2018 data release, the Planck 2018 TT+lowE+lensing at the $95\%$ CL, with the inclusion of the running of the running, provides the following observable bounds \cite{planck}
\begin{align}
n_{s}&=0.9587\pm 0.0112\nonumber\\
\alpha_{s}&=0.013\pm 0.024
\end{align}
Although the superconformal $\alpha$-attractors predict a scalar spectral index of the form $n_{s}=1-2/\Delta N$ as in Eq. \hyperref[eq:ns_alpha]{15}, the actual values of $n_{s}$ calculated for this model are rather low. Following (\hyperref[eq:ns_alpha]{15}), $\Delta N=51.2$ would have yielded $n_{s}\approx0.961$ which is within $2\sigma$ of $n_{s}$ without considering the running of the spectral index. Expressing the amount of observable inflation as $\Delta N=N_{\text{peak}}+\Delta n_{\text{peak}}$, where $\Delta n_{\text{peak}}$ is the e-folds from the peak to the end, a better approximation for the spectral index may be expressed via $n_{s}\approx 1-2/N_{\text{peak}}$. One way to understand this discrepancy is to realize the $\Delta n_{\text{peak}}$ e-folds of inflation after $N_{\text{peak}}$ represents a secondary inflationary stage \cite{dalianis}. The value of $n_{s}$ is thus slightly away from the $2\sigma$ interval. However, it should be noted that $\alpha$-attractor models of this kind can predict $n_{s}$ that is within the $68\%$ C.L., without considering the running, by displacing the inflection point near the end of inflation. Such a modification produces very light PBHs and the existence of DM is explained by their evaporation remnants \cite{Dalianis:2019asr} . On the contrary, the amount of primordial gravitational waves that are actually produced in this model is somewhat larger than predicted by the superconformal $\alpha$-attractors. For $\Delta N=50.6$, the scalar-to-tensor ratio is predicted to be $r\approx 0.00469$ for $\alpha=1$. The value of $r$ calculated using this potential is almost twice as large. Nevertheless, the scalar-to-tensor ratio is within the upper bound established by the BICEP2/Keck observations \cite{BICEP} where $r<0.07$ at $k=\unit[0.05]{Mpc^{-1}}$
\subsection{Implications of reheating}
So far in the analysis, there has been no discussion on the reheating phase. At the end of inflation, the inflaton condensate begins to oscillate around the minimum of the potential and must decay into the Standard Model particles and possibly dark matter. This transition from the super-cooled state at the end of inflation to the hot radiation plasma of the radiation epoch is called reheating. Depending on the decay rate of the inflaton, reheating can be short or prolonged implying large and small reheat temperatures $T_{\text{reh}}$ respectively. It is important to take reheating into account since, up to this point, it has been assumed that the scales that collapse to form PBHs occur during the RD. A prolonged period of reheating implies that it is possible for these scales to re-enter before the onset of RD. In such a case, the mathematical formalisms used to calculate the formation fraction and everything else that followed need to be modified. The reason is that the reheating phase can be thought as being dominated by pressureless matter with $w_{\text{reh}}\simeq 0$, which is essentially an early matter-dominated epoch. Gravitational collapse takes place easily now that there are very little pressure forces to compete against.\\
\indent The method (and notations) discussed in \cite{dalianis} has been used to check whether scales of interest (PBH forming scales) re-enter the horizon during reheating or after. Hence, it is important to establish the epoch of re-entry of such scales. Assuming that a scale $k^{-1}$ exits the horizon $\Delta N_{k}$ e-folds before the end of inflation, upon re-entry
\begin{equation}
\left( \frac{a_{k,\text{re}}}{a_{\text{end}}} \right)^{\frac{1}{2}(1+3w)}=e^{\Delta N_{k}}
\end{equation}
where $a_{k,\text{re}}$ is the scale factor at the time of re-entry. Defining the post-inflationary e-folds leading to the re-entry of $k^{-1}$ as $\tilde{N}_{k}$, we also have
\begin{equation}
\tilde{N}_{k}=\ln\left( \frac{a_{k,\text{re}}}{a_{\text{end}}} \right)
\end{equation}
The two definitions of e-foldings are related to each other via
\begin{equation}\label{eq:eq_50}
\tilde{N}_{k}=\frac{2}{1+3w}\Delta N_{k}
\end{equation}
For example, $\tilde{N}_{k}$ for PBH forming scales can be easily calculated by determining $\Delta N_{k,\text{peak}}$ from the curvature power spectrum. The value of $\tilde{N}_{k}$ naturally creates the delineation between the scales those enter during reheating and the ones during RD. Assuming that the scale factor at the end of reheating is given by $a_{\text{reh}}$, then the number of e-folds during this period is $\tilde{N}_{\text{reh}}=\ln\left( a_{\text{reh}}/a_{\text{end}} \right)$. We note that the energy density at the end of inflation is given by $\rho_{\text{end}}=3H_{\text{end}}^{2}M_{p}^2$, which can be used to relate the two phases through $\rho_{\text{reh}}=\rho_{\text{end}}e^{-3N_{\text{reh}}(1+w_{\text{reh}})}$. Then,
\begin{equation}\label{eq:nktilde}
\tilde{N}_{k}=\frac{1}{3(1+w_{\text{reh}})}\ln\left( \frac{\rho_{\text{end}}}{\rho_{\text{reh}}} \right)
\end{equation}
If $\tilde{N}_{k}>\tilde{N}_{\text{reh}}$, the scale $k^{-1}$ becomes subhorizon during the radiation era. On the other hand, $\tilde{N}_{k}<\tilde{N}_{\text{reh}}$ represents scales that re-enter during the reheating phase. The amount of observable inflation also crucially depends on the post-inflationary expansion. Importantly, $\Delta N$ is related to $\tilde{N}_{k}$ through the following equation (assuming $w_{\text{reh}}\simeq 0$)
\begin{equation}\label{eq:4.16}
\Delta N\simeq 57.3+\frac{1}{4}\ln\left( \frac{\epsilon_{\star}V_{\star}}{\rho_{\text{end}}} \right)-\frac{1}{4}\tilde{N}_{\text{reh}}
\end{equation}
Although a criterion has been established that tells us whether or not a particular scale re-enters during reheating, it can be further refined by placing limits on $\Delta N$. Using Eq. \hyperref[eq:eq_50]{50}, $\Delta n_{\text{peak}}=\tilde{n}_{\text{peak}}/2$ where $\tilde{n}_{\text{peak}}$ is the number of e-folds after the end of inflation when $k^{-1}_{\text{peak}}$ became superhorizon. Considering a peak which re-enters the horizon by the time the reheating phase just ends, then $\Delta n_{\text{peak,c}}=\tilde{N}_{\text{reh}}/2$. In such a case, the amount of observable inflation becomes $\Delta N=N_{\text{peak}}+\tilde{N}_{\text{reh}}/2$. Plugging this into Eq. \hyperref[eq:4.16]{52}, this threshold e-folding can be expressed in terms of $N_{\text{peak}}$.
\begin{equation}\label{eq:4.17}
\Delta n_{\text{peak,c}}=\frac{2}{3}\left[ 57.3 +\frac{1}{4}\ln\left( \frac{\epsilon_{\star}V_{\star}}{\rho_{\text{end}}} \right)-N_{\text{peak}} \right]
\end{equation}

\begin{figure}[t]
\centering

\subfloat{\includegraphics[scale=0.5]{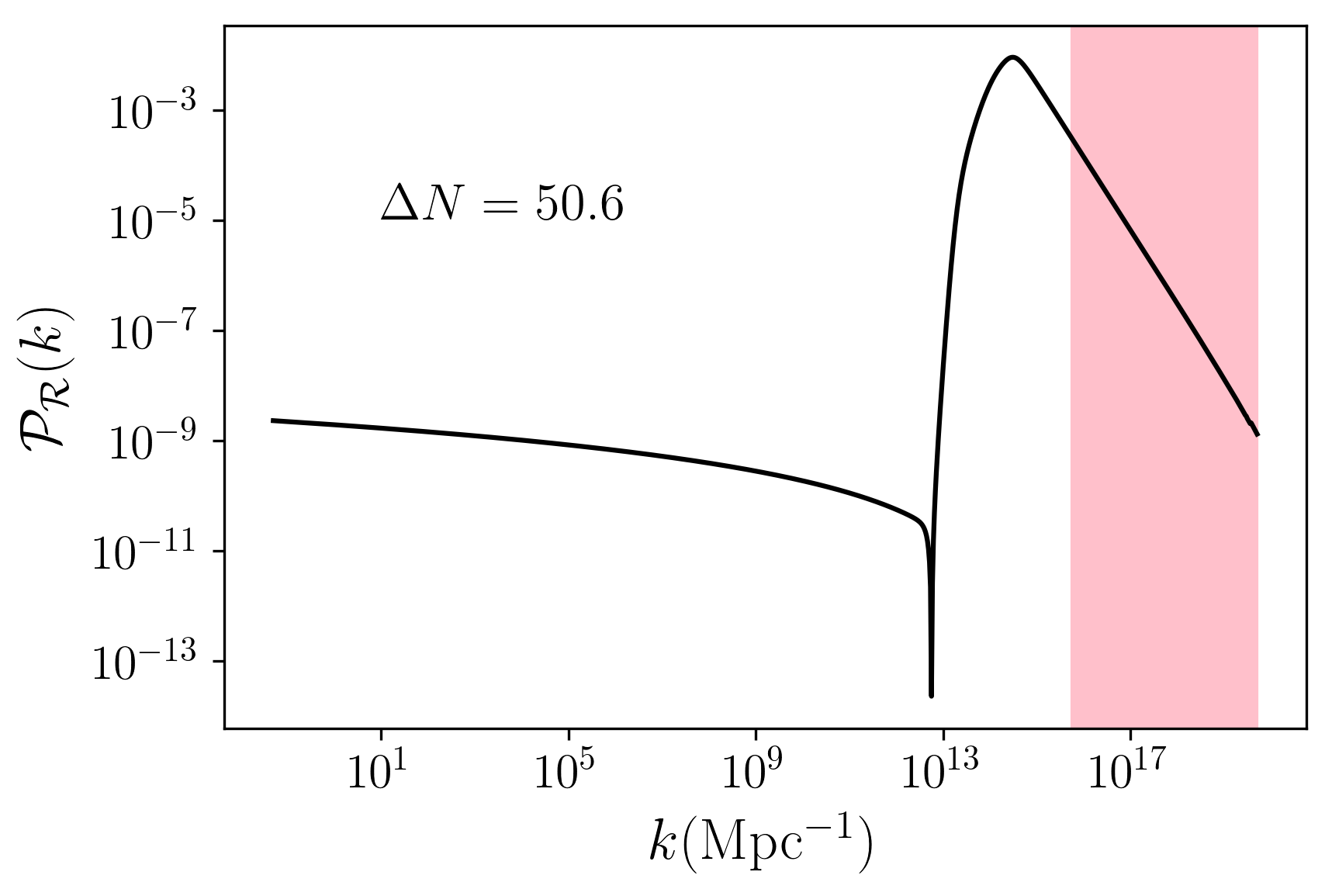}}
\subfloat{\includegraphics[scale=0.5]{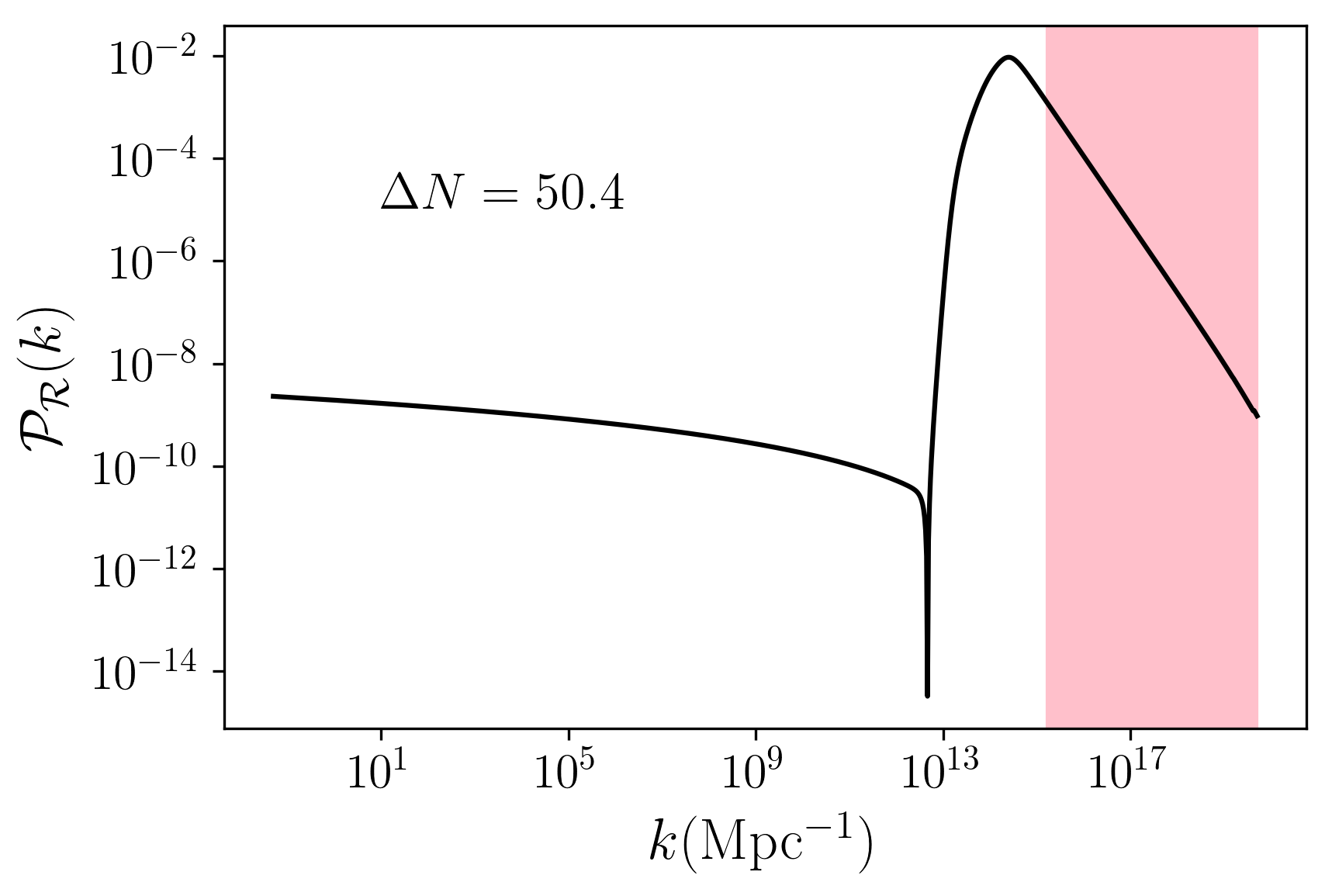}}

\vfill

\subfloat{\includegraphics[scale=0.5]{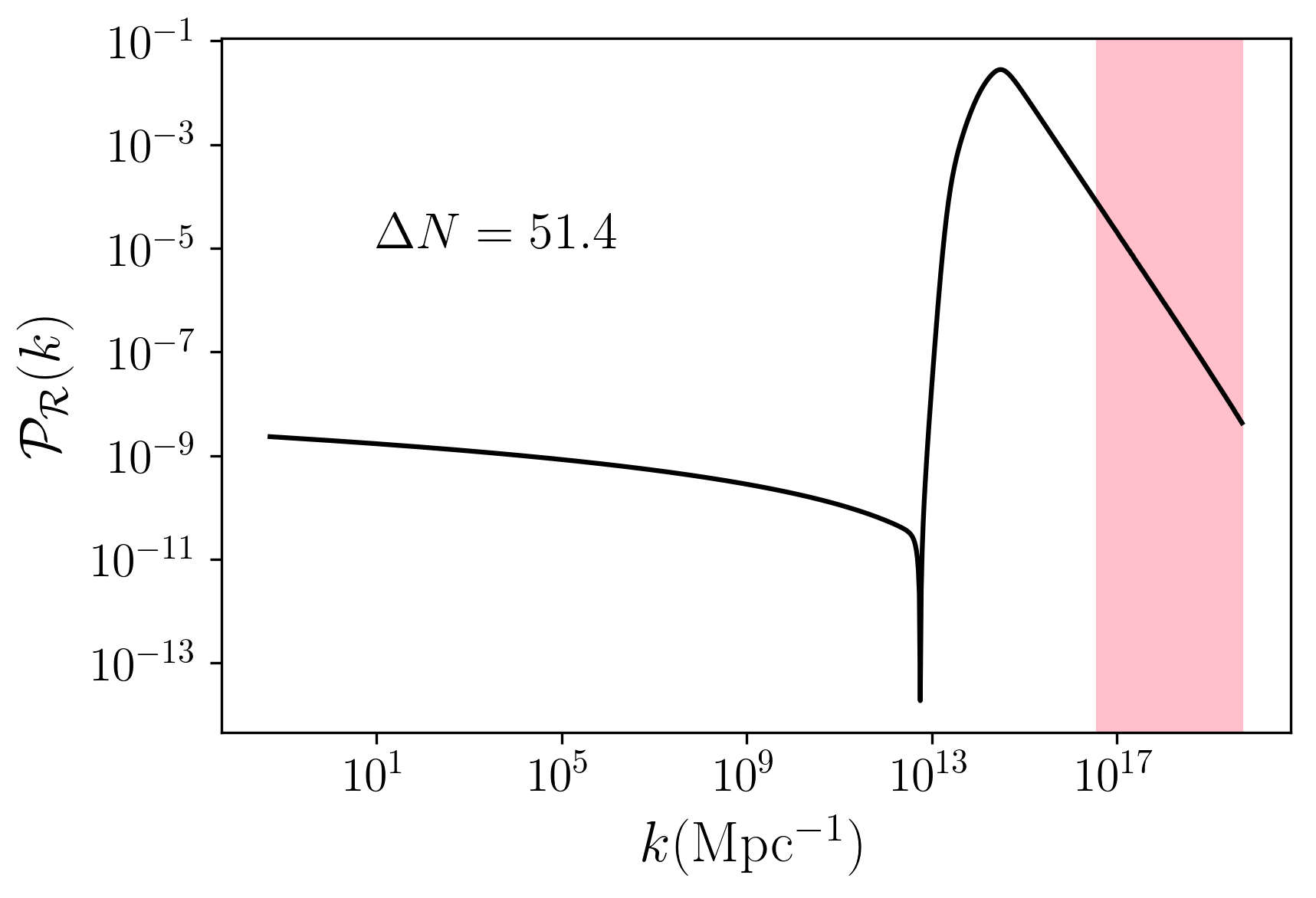}}
\subfloat{\includegraphics[scale=0.5]{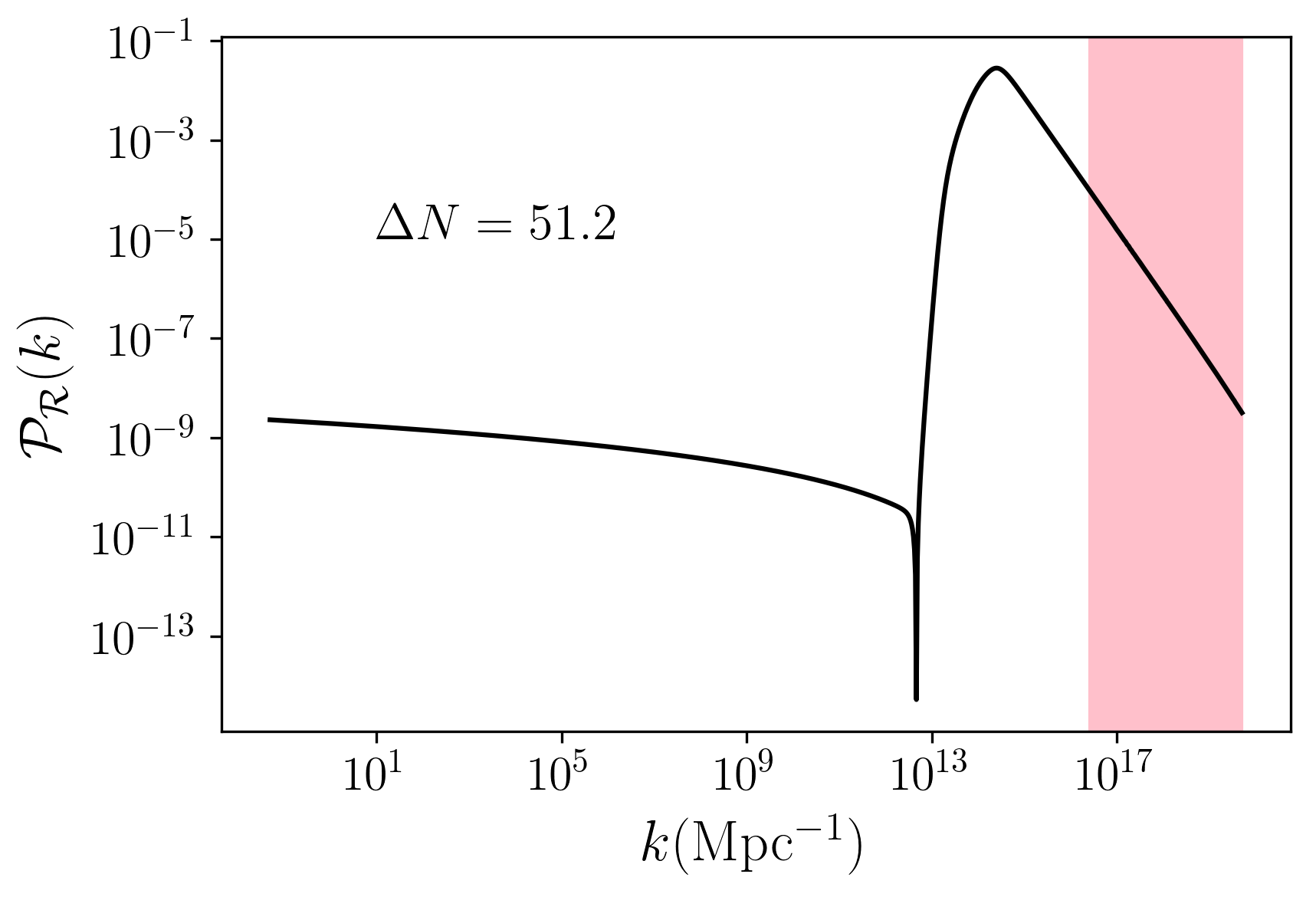}}


\label{fig:figure7}
\caption{Summary of the curvature power spectra for parameter sets 1 (top panel) and 2 (bottom panel). The shaded regions illustrate the scales that re-enter the horizon during the reheating phase.}
\end{figure}
To improve this cut-off, the expression in Eq. \hyperref[eq:4.17]{53} can be simplified further. With the inflaton potential considered in this work, both parameter sets produce very similar background evolution and power spectra with the exception being in \#2 inflation lasts slightly longer and $\mathcal{P}_{\mathcal{R}}^{\text{max}}$ is higher. For both parameter sets $\rho_{\text{end}}\sim 10^{-11}M_{p}^{4}$ and hence $\ln(\epsilon_{\star}V_{\star}/\rho_{\text{end}})^{1/4}\approx -1.3$. Then,
\begin{equation}
\Delta n_{\text{peak},c}\simeq\frac{2}{3}\left( 56-N_{\text{peak}} \right) 
\end{equation}
For both parameter sets considered here, very similar values for $k_{\text{peak}}$ and $N_{\text{peak}}$ have been calculated, as summarized in table \hyperref[Tab:table4]{IV}.\\
\begin{table}[h]
\centering
\begin{tabular}{|l|l|l|l|l|l|}
\hline
\#                      & $\Delta N$                & $k_{\text{peak}}$                         & $N_{\text{peak}}$         & $\Delta n_{\text{peak}}$   & $\tilde{N}_{\text{reh}}$  \\ \hline
1 & 50.6 & $3.09\times 10^{14}$ & 36.4 & 13.07 & 21.6 \\ \hline
1 & 50.4 & $2.43\times 10^{14}$ & 36.1 & 13.3  & 22.4 \\ \hline
2 & 51.4 & $3.09\times 10^{14}$ & 36.4 & 13.07 & 18.4 \\ \hline
2 & 51.2 & $2.43\times 10^{14}$ & 36.1 & 13.3  & 19.2 \\ \hline
\end{tabular}
\label{Tab:table4}
\caption{Summary of $N_{\text{peak}}$ and $\Delta n_{\text{peak}}$ for both parameter sets.}
\end{table}
\indent Using information from table \hyperref[Tab:table4]{IV}, the cut-off e-foldings such that the peak in the curvature power spectra re-enter the horizon after the reheating phase is $\Delta N>49.4$. Hence, for both sets of parameters and $\Delta N$ that have been considered, $k^{-1}_{\text{peak}}$ become superhorizon during the radiation epoch. Although in Fig. \hyperref[fig:figure6]{6} the $f_{\text{PBH}}(M)$ curves only cover a small mass range and, by extension, a small range in $k$, it still needs to be determine whether or not all the relevant, PBH forming scales that re-enter the horizon before and after $k^{-1}_{\text{peak}}$ do so after reheating. This is because for both parameter sets, $\tilde{N}_{\text{reh}}$ is rather prolonged. To check this, the values for $\tilde{N}_{\text{reh}}$ can be used from table \hyperref[Tab:table4]{IV} to calculate the scales $k_{\text{reh}}$ that entered the horizon just at the end of reheating. This information can then be used, along with Eq. \hyperref[eq:pbh_mass]{37}, to determine the mass of PBH such a scale can give rise to upon collapse $M(k_{\text{reh}})$. In such a case all $M(k)<M(k_{\text{reh}})$ will form in the reheating phase. This has been illustrated in Fig. \hyperref[fig:figure7]{7}. The shaded areas represent all the scales that re-enter the horizon during the reheating phase. Since the bulk of the PBH forming scales are the ones in the vicinity of $k^{-1}_{\text{peak}}$, it is not surprising that the scales that become subhorizon during reheating do not play an important role. For example, in parameter set 1 with $\Delta N = 50.4$, all scales smaller than $\unit[1.68\times 10^{15}]{Mpc^{-1}}$ re-enter during reheating. As a result, the maximum mass of PBHs that these can form will be of the order $\unit[10^{15}]{g}$. Since $k_{\text{reh}}$ for all the other $\Delta N$ are higher, the masses of PBHs that can possibly form during reheating are lower and should not affect the $f_{\text{PBH}}$ curves shown in Fig. \hyperref[fig:figure6]{6}.

\subsection{Observational constraints}\label{sec:section4D}
\begin{figure}[t]
\centering
\includegraphics[scale=1]{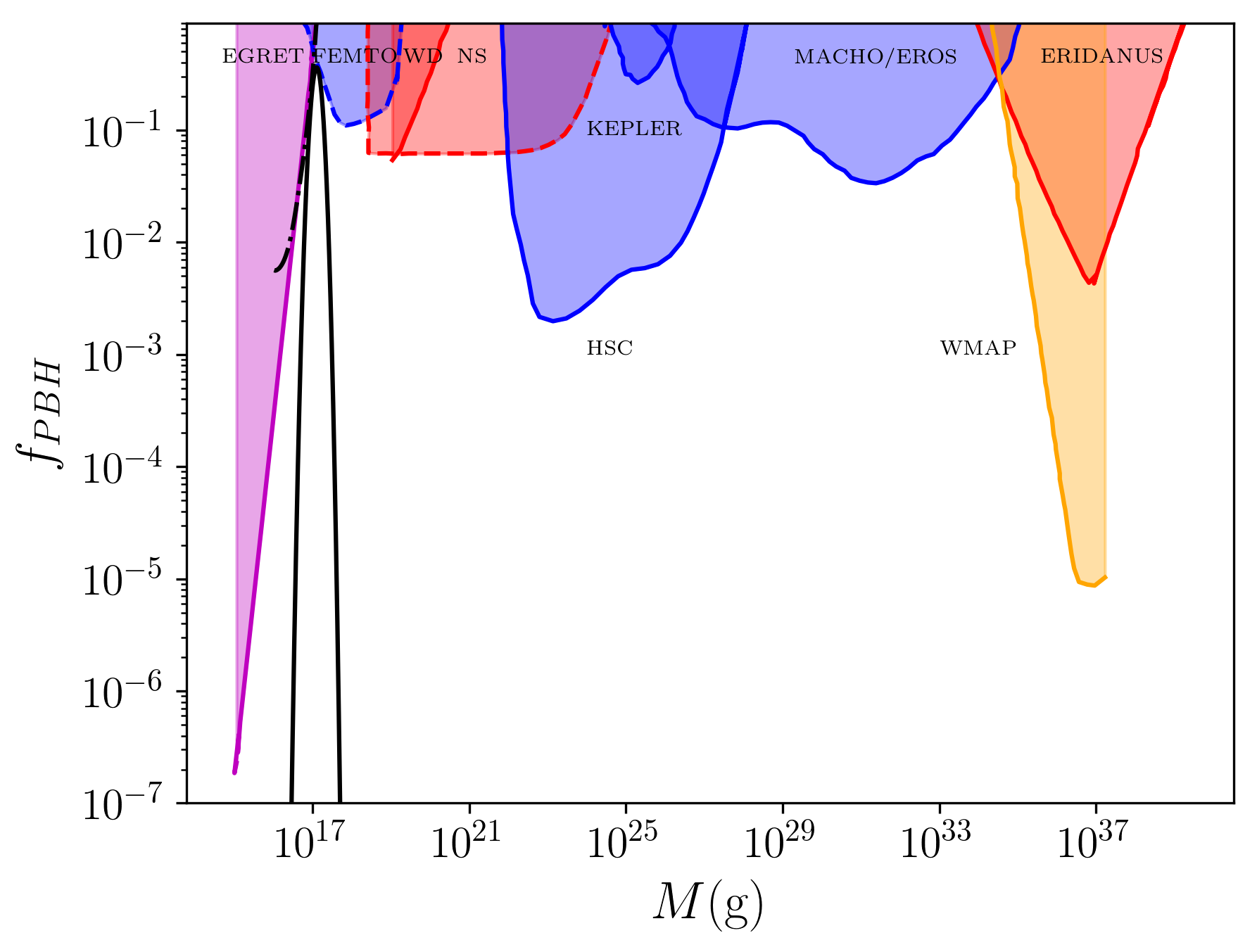}
\caption{Constraints on PBH abundance in the mass range $10^{-18}M_{\odot}-10^{3}M_{\odot}$. The different categories include: (i) evaporation (extragalactic $\gamma$-ray background \cite{carr2016constraints} in \textit{magenta} and $e^{\pm}$ annihilation and $\unit[511]{keV}$ gamma-ray line \cite{Laha:2019ssq} in black dot-dashed), (ii) gravitational lensing: femtolensing of $\gamma$-ray bursts \cite{katz2018femtolensing}, microlensing of the Magellanic Clouds by MACHO and EROS \cite{tisserand2007} and microlensing from Subaru \cite{subaru} in \textit{blue} (iii) dynamical constraints from white dwarves (WD) \cite{graham2015dark}, neutron star (NS) \cite{capela2013constraints} capture and survival of a star cluster in Eridanus II (E) \cite{brandt2016constraints} in \textit{red} and (iv) accretion constraints \cite{poulin2017cmb} in \textit{orange}. The black solid curve represents a result from parameter set 2. It should be noted that the constraints shown in dashed lines are ones over which some disputes exist. Figure adapted from \cite{carr2,dalianis}}.
\label{fig:figure8}
\end{figure}

Although it is theoretically possible to devise a scenario where PBHs form all of the dark matter, in reality the likelihood of such an event is highly constrained by observations. In Fig. \hyperref[fig:figure6]{6}, we see a few examples of observational constraints that are relevant for low mass PBHs. The constraints on $f_{\text{PBH}}(M)$ span a very large range of masses and there are only a few windows which inflation models can exploit to form PBHs that can comprise at least a sizeable proportion of dark matter. Constraints on $f_{\text{PBH}}(M)$ can be broadly divided into \cite{sasaki,carr2} (i) evaporation (ii) gravitational lensing (iii) dynamical and (iv) accretion constraints with others which are more indirect (Fig. \hyperref[fig:figure8]{8} contains a detailed plot of these different constraints).\\
\indent The evaporation constraints \cite{carr2010} are most relevant for PBHs with $M\leq \unit[10^{16}]{g}$ since these are the ones which are expected to have evaporated by Hawking radiation (or ones which are evaporating). This constaint is shown in the graph labelled as "EGRET" and we see that it get progressively weaker for larger masses. On the other end of the mass spectrum, PBH abundance is severely constrained by a lack of observation of accretion effects on the CMB \cite{poulin2017cmb}; as a result $M\geq 10^{2}M_{\odot}$ cannot constitute an important fraction of dark matter. PBH dark matter is also constrained around the solar mass area through the survival of a star cluster in the dwarf galaxy Eridanus II \cite{brandt2016constraints}. The masses in between are also constrained- although considerably weakly in some areas.\\
\indent There are some mass windows in which PBHs could constitute a large fraction of dark matter. From around $\unit[10^{17}]{g}-\unit[10^{21}]{g}$ (or $10^{-16}-10^{-12}M_{\odot}$), the constraints from femtolensing \cite{katz2018femtolensing}, white dwarf and neutron star capture \cite{graham2015dark,capela2013constraints} are considerably less stringent than other areas and primordial black holes could account for around 10-20\% of the dark matter \cite{ballesteros}. Specifically, the femtolensing and neutron star data have been shown using dashed lines, indicating that there are uncertainties. The neutron star survival data come from globular clusters and is sensitive to the dark matter density taken there ($>\unit[10^3]{GeV\;cm^{-3}}$ \cite{capela2013constraints}). According to \cite{carr2}, the dark matter density in globular clusters is much lower. Femtolensing of gamma-ray bursts (GRBs) also places constraints on $f_{\text{PBH}}$ in that region. However, it was pointed out that, due to their large sizes, GRBs would not be the best candidates for femtolensing searches \cite{katz2018femtolensing}. Another constraint of importance arises from the survivability of white dwarves (WD). PBHs passing through white dwarves can cause localized heating due to dynamical friction and induce thermonuclear runaway. Hence, observing white dwarves in a certain mass range puts constaints on PBH aboundance. A recent study \cite{Montero-Camacho:2019jte} concluded that the minimum mass required for a PBH to induce thermonuclear runaway is around $\unit[10^{21}]{g}$. All of these combined leave a mass window which is potentially unconstrained and open for future investigation. Another recent study \cite{Laha:2019ssq} established new constraints for low mass PBHs arising from $e^{\pm}$-pairs produced from PBH evaporation and subsequent annihilation in the Galactic Center (using the \unit[511]{keV} gamma ray line). The constraints were established using isothermal and Navarro-Frenk-White (NFW) profiles. This is shown in Fig. \hyperref[fig:figure8]{8} as the black dashed-dotted line along EGRET.\\
\indent Finally, it should be emphasized that the $f_{\text{PBH}}$ constraints shown in Fig. \hyperref[fig:figure8]{8} apply for PBHs with a monochromatic mass function. In reality, however, the expectation is that PBHs will be described by extended mass functions for which which the constraints become more stringent. A thorough analysis of this can be found in \cite{Carr:2017jsz} where the monochromatic $f_{\text{PBH}}$ has been compared with lognormal and power law mass functions.

\section{Second order gravitational waves}
At linear order in perturbation theory, the Fourier modes of the scalar and tensor perturbations are decoupled and evolve independently of each other. The evolution of the tensor modes is governed by 
\begin{equation}
v_{\bm{k}}''+\left( k^{2}-\frac{a''}{a} \right)v_{\bm{k}}=0    
\end{equation}
which is basically the Mukhanov-Sasaki equation. However, if second order effects are considered, it has been established that the evolution of tensor perturbations is sourced by the first order scalar perturbations (although there is no mixing among the second order scalar, vector and tensor quantities). Due to the fact that the inflationary curvature power spectrum is largely undetermined at small scales, detection of second order GWs would be a unique way of constraining the primordial scalar power spectrum \cite{baumann2007gravitational}. Moreover, they would indirectly point toward the possibility of PBH formation from models that produce amplified $\mathcal{P}_{\mathcal{R}}$ at small scales. At second order in perturbation theory, the metric is written as 
\begin{equation}
ds^{2}=a^{2}(\tau)\left[ -\left(1+2\Phi^{(1)}+2\Phi^{(2)}  \right)d\tau^{2}+2V_{i}^{(2)}d\tau dx^{i}+\left\{ \left( 1-\Psi^{(1)}-2\Psi^{(2)} \right)\delta_{ij}+\frac{1}{2}h_{ij} \right\}dx^{i}dx^{j} \right]
\end{equation}
where $h_{ij}$ simply refers to the second order tensor perturbations. The evolution of the tensor perturbation of the Fourier modes is now expressed by the following inhomogeneous ODE
\begin{equation}
h_{\bm{k}}''(\tau)+2aHh'_{\bm{k}}(\tau)+k^{2}h_{\bm{k}}(\tau)=S_{\bm{k}}(\tau)
\end{equation}
with the following source term (the Fourier components of the transverse-tracefree source term)
\begin{align}
S_{\bm{k}}(\tau)&=\int\frac{d^{3}\bm{q}}{(2\pi)^{3/2}}\bm{q}\;^{2}\left[ 1-\left( \frac{\bm{k}\cdot\bm{q}}{|\bm{k}||\bm{q}|} \right)^2 \right]\Bigg[ 12\Phi_{\bm{k}-\bm{q}}(\tau) \Phi_{\bm{k}}(\tau)  \nonumber\\
&\quad{} +8\left(\tau\Phi_{\bm{k}-\bm{q}}(\tau)+\frac{\tau^2}{2}\Phi_{\bm{k}-\bm{q}}'(\tau)  \right)\Phi_{\bm{q}}'(\tau)\Bigg]
\end{align}
The approach is similar to what had been done in \cite{baumann2007gravitational} in their analytical calculation where anisotropic stresses were ignored such that $\Psi = \Phi$. The Bardeen potential is related to the curvature perturbation through $\Phi=2\mathcal{R}_{k}/3$. Since the curvature perturbation is enhanced in the ultra slow-roll region, we expect the source term to also get enhanced for scales which exit the horizon during that time. Much like in the case of the Mukhanov-Sasaki equation, a field redefinition $v_{\bm{k}}=ah_{\bm{k}}$ is applied to obtain the following
\begin{equation}
v_{\bm{k}}''+\left( k^2 -\frac{a''}{a} \right)v_{\bm{k}}=aS_{\bm{k}}(\tau)
\end{equation}
This is an inhomogeneous ordinary differential equation and its formal solution may be expressed in terms of a Green function
\begin{equation}
h_{\bm{k}}(\tau)=\frac{1}{a(\tau)}\int d\tau'G_{\bm{k}}(\tau,\tau')a(\tau')S_{\bm{k}}(\tau')
\end{equation}
With the solutions, the spectral energy density of the gravitational waves\footnote{$\Omega_{\text{GW}}$ is the fraction of critical energy density in gravitational waves per logarithmic interval of frequency.} can be computed.

\[\Omega^{(2)}_{\text{GW}}(k,\tau)=A^{(2)}_{\text{GW}}\mathcal{P}_{\mathcal{R}}^{2} \begin{cases} 
      
      \frac{a(\tau)}{a_{\text{eq}}}\frac{k}{k_{\text{eq}}} &\text{for } k<k_{\text{eq}} \\
      \frac{a(\tau)}{a_{\text{eq}}}\left(\frac{k}{k_{\text{eq}}}\right)^{2-2\gamma} &\text{for } k_{\text{eq}}<k<k_{c}(\tau) \\
      \frac{a_{\text{eq}}}{a(\tau)} &\text{for } k>k_{c}(\tau) 
   \end{cases}
\]

\begin{figure}[t]
    \centering
    \includegraphics[scale=0.8]{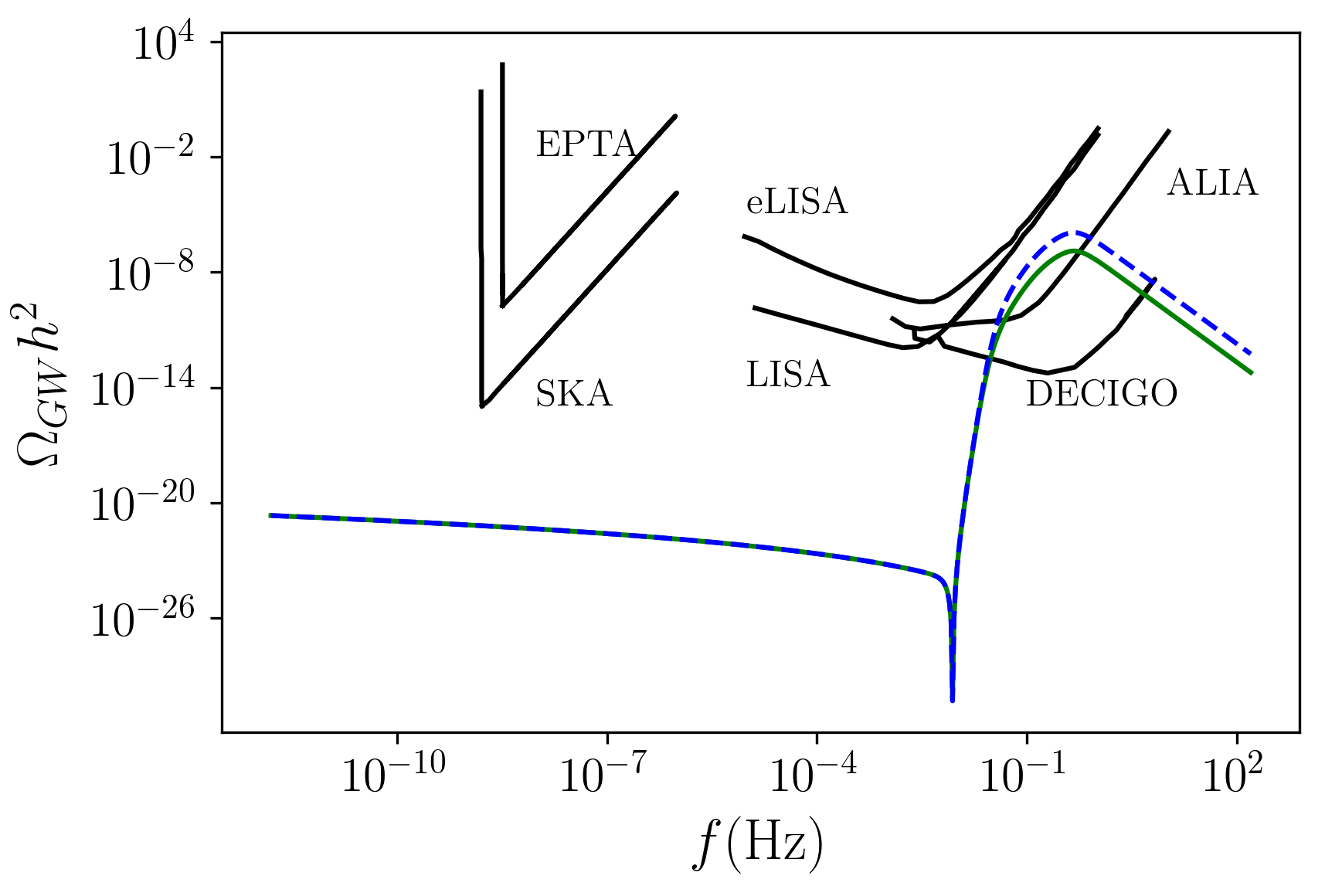}
    \caption{Energy density of second order gravitational waves as a function of frequency. The green (parameter set 1) and blue-dotted (parameter set 2) represent $\Omega_{\text{GW}}h^2$ predicted by the inflation model. The solid black lines represent the experimental sensitivities of some current and proposed GW detectors. The sensitivity curves have been adapted from \cite{Moore_2014}.}
    \label{fig:figure9}
\end{figure}
\begin{figure}\label{fig:figure12}
\centering
\includegraphics[scale=0.55]{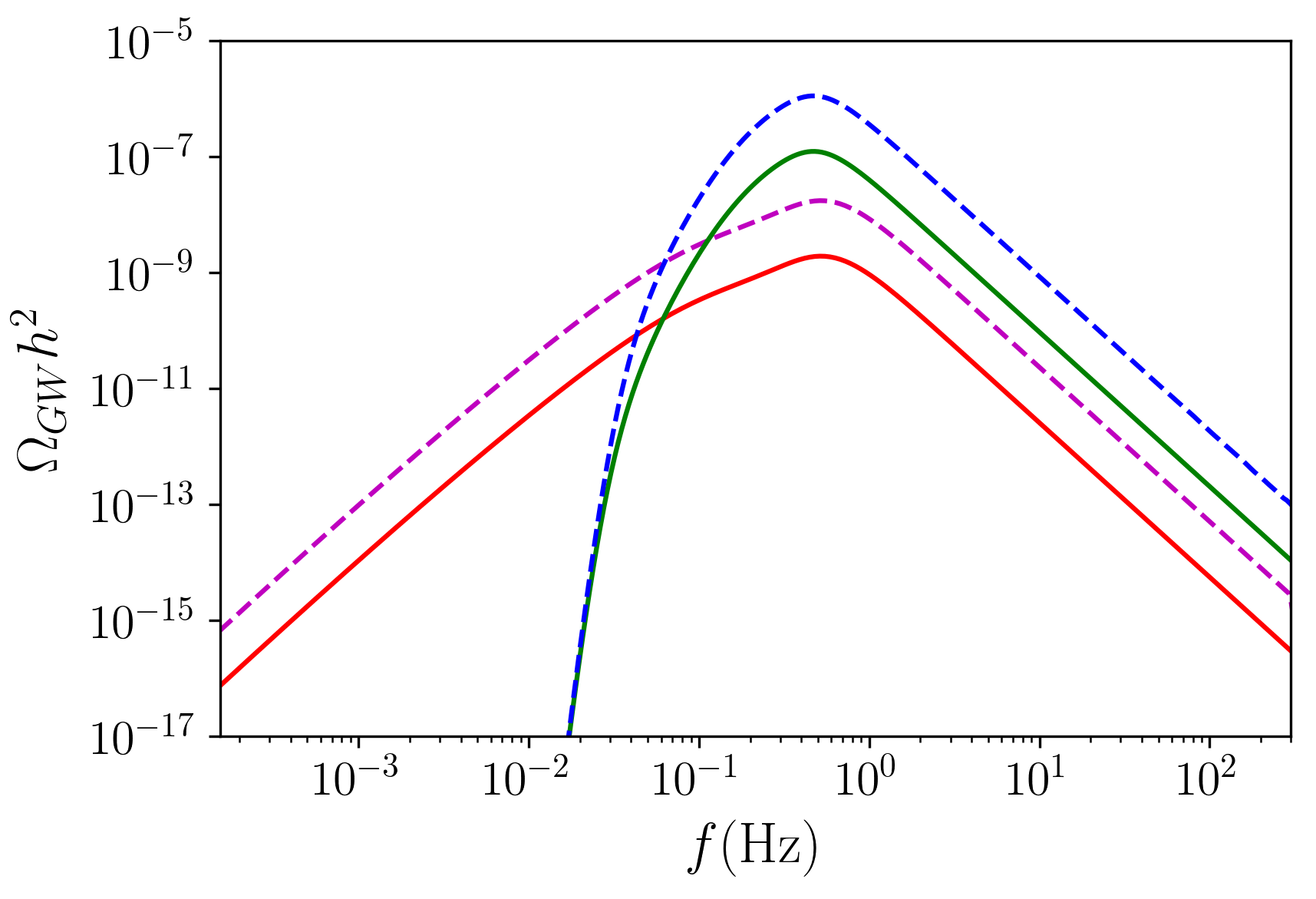}
\includegraphics[scale=0.55]{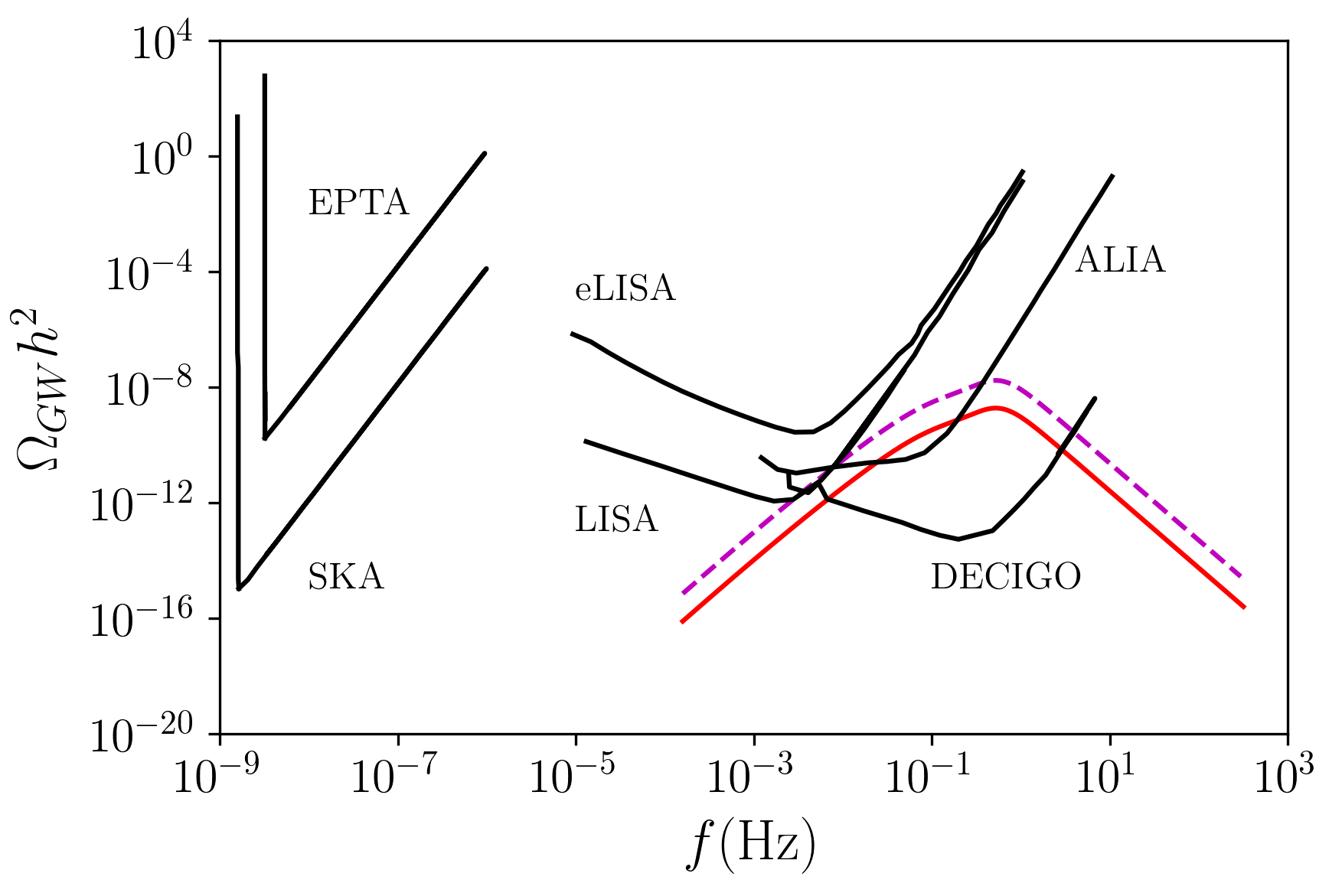}
\caption{\textit{Left}: A comparison of the GW energy density spectrum using the two approaches. The solid red and dotted magenta lines represent parameter sets 1 and 2 obtained from Eq. \hyperref[eq:equation63]{63}. \textit{Right}: Overall comparison with GW sensitivity curves.}
\end{figure}

\noindent where $k_{c}(\tau)=(a/a_{\text{eq}})^{1/(1-\gamma)}k_{\text{eq}}$, $A^{(2)}_{\text{GW}}\approx 10$ and $\gamma\approx 3$ \cite{baumann2007gravitational}. At matter-radiation equality, the comoving wavenumber is $k_{\text{eq}}=\unit[0.07\Omega_{\text{m}}h^{2}]{Mpc^{-1}}\approx \unit[0.01]{Mpc^{-1}}$. Since the redshift at matter-radiation equality is known, it can be used to compute the scale factor (assuming the current scale factor is set to unity). For the model presented here, the amplification in the curvature power spectrum occurs for modes around $\unit[10^{14}]{Mpc^{-1}}\gg k_{\text{eq}}$. They also happen to be larger than $k_{c}$ at present time $\tau_{0}$. Then,
\begin{equation}
\Omega_{\text{GW}}^{(2)}(k,\tau_{0})\simeq 10\mathcal{P}_{\mathcal{R}}^{2}a_{\text{eq}}
\end{equation}
The energy density of gravitational waves over the critical density of the universe has been plotted in Fig. \hyperref[fig:figure9]{9} as a function of frequency. The energy spectrum of the second order GWs produced by this model lies within the sensitivities of ALIA (Advanced Laser Interferometer Antenna) and DECIGO (Deci-hertz Interferometer Gravitational wave Observatory) which are future planned detectors. It should be noted that, while this method produces interesting results, the expressions used were derived under numerous simplifications. A semi-analytical expression for the second order GW power spectrum $\mathcal{P}_{h}$, and hence the energy density spectrum, was derived in \cite{Kohri:2018awv} using fewer simplifications and it would be useful to compare the results obtained from both approaches. The exact expression for the GW energy density is
\begin{equation}
\Omega_{\text{GW}}(k,\tau)=\frac{1}{24}\left( \frac{k}{\mathcal{H}(\tau)} \right)^{2}\overline{\mathcal{P}_{h}(k,\tau)}
\end{equation}\label{eq:equation63}
where the overbar indicates average over conformal time. The conformal time averaged power spectrum of the tensor perturbations is then given by
\begin{equation}\label{eq:equation63}
\overline{\mathcal{P}_{h}(k,\tau)}=2\int_{0}^{\infty}dt\int_{-1}^{1}ds\left[ \frac{t(2+t)(s^2 -1)}{(1-s+t)(1+s+t)} \right]^2\overline{I(v,u,x)_{\text{RD}}^{2}}\mathcal{P}_{\mathcal{R}}(vk)\mathcal{P}_{\mathcal{R}}(uk)
\end{equation}
where $u=(t+s+1)/2$, $v=(t-s+1)/2$ and $x=k\tau$. The function $\overline{I^{2}_{\text{RD}}(v,u,x)}$ is expressed in terms of a rather complicated combination of terms, the exact form of which can be found in \cite{Kohri:2018awv,inomata2019gravitational}. Although the integral can be simplified if the $\mathcal{P}_{\mathcal{R}}(k)$ are modelled using lognormal distributions, the second order tensor power spectrum is computed exactly using the curvature power spectra data obtained by solving the Mukhanov-Sasaki equation. It is seen from Eq. \hyperref[eq:equation63]{63} a very important difference between the two approaches arises from the fact that the curvature power specta are integrated over, which immediately hints that there should be quantitative deviations between the two. During the radiation epoch, the second order GWs are produced around horizon entry without any significant contributions afterwards since the transfer function $\Phi_{\bm{k}}$ decays during this epoch. To obtain the energy density spectrum, Eq. \hyperref[eq:equation63]{63} is integrated from $k=10^{11}$ to $\unit[10^{18}]{Mpc^{-1}}$ (which is the interesting range of wavenumbers for which $\Omega_{\text{GW}}h^{2}$ is expected to lie within the sensitivities of future detectors) to obtain $\Omega_{\text{GW}}^{(2)}h^2$ at the time of formation. The present day value is then obtained by accounting for redshift \cite{inomata2019gravitational,Bartolo:2018rku}.
\begin{equation}
\Omega_{\text{GW}}^{(2)}(k,\tau_{0})=0.83\left( \frac{g_{\star}(\tau_{\text{form}})}{10.75} \right)^{-1/3}\Omega_{\text{r},0}\Omega_{\text{GW}}^{(2)}(k,\tau_{\text{form}})
\end{equation}  
where $g_{\star}$ represents the effective number of relativistic degrees during the epoch GW generation and $\Omega_{\text{r},0}h^{2}=4.2\times 10^{-5}$ is the radiation energy density of the present day. Figure \hyperref[fig:figure10]{10} shows a comparison of the two two results. Although the $\Omega_{\text{GW}}^{(2)}$ curves peak around the same $k$ value for both computations, the second approach produces a wider curve with a peak atleast a couple of orders of magnitude lower than the approximation. This, however, does not affect the possibilities of detection as the curves are still within the sensitivities of ALIA and DECIGO.\footnote{In fact, it can be seen from Fig. \hyperref[fig:figure12]{12} that LISA might be sensitive enough to detect them for parameter set 2.} 
\label{fig:figure10}
\section{Conclusions}
In this work, a PBH formation scenario, through a single field inflation model with a quasi-inflection point, was examined. A potential was constructed out of the inflationary $\alpha$-attractor framework which possessed a plateau-like region, providing appropriate conditions for an ultra slow-roll regime. During such a transitory phase, the curvature perturbations were amplified which re-entered the horizon after inflation and collapsed to form PBHs. Two parameter sets were studied out of which one of them produced PBHs from overdensities $\delta_{c}\sim 0.4$ which were close to the numerically favoured threshold overdensity. The mass distributions of PBHs produced from both sets of parameters were centered around $\unit[10^{17}]{g}$. With both parameters and the different e-fold runs, it was found that this model can produce PBHs that can comprise more than 20\% of the dark matter. Although it was mentioned in Sec. \hyperref[sec:section4D]{IV.D} that the mass range relevent to this work is potentially unconstrained, the $f_{\text{PBH}}$ was calculated respecting those bounds. A scheme was devised to also verify whether or not PBH forming scales $k_{\text{peak}}\sim\unit[10^{14}]{Mpc^{-1}}$ re-entered the horizon during reheating- which was shown to the contrary. Finally, the energy density spectrum of second order gravitational waves was calculated, which was shown to lie within the sensitivities of future GW detectors.\\
\indent As mentioned earlier, one of the apparent shortcomings of such models is the high level of fine-tuning required to produced the desired results. Although not a problem in itself, it does allude to the fact that PBHs do not arise naturally in such cases. Another issue is the relatively low value of $n_{s}$ predicted by the model. The pivot scale could be set such that the value $n_{s}$ is much improved. However, that would mean that $\mathcal{P}_{\mathcal{R}}$ would have to peak at higher values of $k$, meaning lower mass PBHs.\\
\indent Further work can be done in specific areas to supplement the conclusions of this work. In the study of PBH formation, the mass fraction is usually calculated assuming Gaussian statistics of the primordial power spectrum. While the primordial power spectrum is very nearly Gaussian, the CMB does reveal small amounts of non-Gaussianities. Primordial non-Gaussianities can have significant impact on the formation fraction since the tail of the distribution can be modified in such cases, leading to a greater fraction of PBH formation. As a result, the parameters presented in this work can actually overproduce PBHs. Moreover, one needs to consider the effects of quantum diffusion, especially in the non-attractor regime of USR, for a more accurate description. This is due to that fact that the usual description of inflation deals with the inflaton field classically. However, subhorizon quantum modes can influence the superhorizon classical trajectory which is neatly described in the programme of stochastic inflation \cite{Biagetti:2018pjj,Pattison:2019hef}. It can be shown that the ratio of the amplitude of the stochastic corrections to that of the classical trajectory is roughly $\sqrt{\mathcal{P}_{\mathcal{R}}}$. For the most part, this is extremely small due to the nature of the primordial power spectrum. In the case of a USR phase, the contributions arising from stochastic corrections can become significant. When quantum diffusion effects become important, it can amplify the curvature perturbation power spectrum in cases which would not have otherwise experienced such an increase \cite{Ezquiaga:2018gbw}. In such an event, both parameter sets that have been used here might actually overproduce PBHs. However, it does mean that other sets in the parameter space become available for investigation. These areas, along with their implications, will be worked on in a future project.
\section{Acknowledgements}
The author is grateful to J. Kapusta for comments on the manuscript. The author also thanks T. Terada regarding email communications which led to a more accurate calculation of the second order GW energy density spectrum.
\section{Note added}
In the original manuscript, the second order GW energy density spectrum was calculated only using the approximation derived in \cite{baumann2007gravitational}. It was later pointed out that a more accurate approach exists \cite{Kohri:2018awv}. The section on observational constaints has also been amended to take into account updated constraints in the relevent mass window.


\bibliography{pbh_mahbub}

\end{document}